\documentclass{article}
 
\begin{document}

\thispagestyle{plain}
\def\upcirc#1{\vbox{\ialign{##cr
$\circ\!$cr \noalign{\kern-0.1 pt\nointerlineskip}
$\hfil\displaystyle{#1}\hfil$cr}}}
\def\X{\bf X}
\def\thorn{\hbox{\rm I}\kern-0.32em\raise0.35ex\hbox{\it o}}
\def\edth{\hbox{$\partial$\kern-0.25em\raise0.6ex\hbox{\rm\char'40}}}
\def\edthbar{\overline{\edth}}
\def\taub{\overline{\tau}}
\def\obar{\overline{o}}
\def\iotabar{\overline{\iota}}
\def\thornprime{\thorn^\prime}
\def\edthprime{\edth^\prime}
\def\i{\iota}
\def\P{{\bf \Psi}}
\def\Thorn{{\bf \thorn}}
\def\Fi{{\bf \Phi}}
\def\bfi{\bf \phi}
\def\bnab{{\bf \nabla}}
\def\L{{\bf \Lambda}}
\def\bxi{{\bf \Xi}}
\def\bep{{\bf \epsilon}}
\def\bdelta{\mbox{\boldmath $\delta$}}
\def\bDelta{{\bf \Delta}}
\def\thornp{{\thorn^\prime}}
\def\edthp{{\edth^\prime}}
\def\bfeta{{\bf \eta}}
\def\pmb#1{\setbox0=\hbox{#1} \kern-.025em\copy0\kern-\wd0
\kern.05em\copy0\kern-\wd0 \kern-.025em\raise.0233em\box0 }
\def\bedth{{\pmb{\edth}}}
\def\bthorn{{\pmb \thorn}}
\def\thornp{{\thorn^\prime}}
\def\edthp{{\edth^\prime}}
\def\bfeta{{\bf \eta}}
\def\bthornp{{\bthorn^\prime}}
\def\bedthp{{\bedth^\prime}}
\def\alemf{{\nabla^{f\plf}\nabla_{f\plf}}}
\def\Ib{\overline{\bf I}}
\def\I{\bf I}
\def\ei{{\rm i}}
\def\Ph{\nthorn}
\def\D{\nedth}
\parindent=0pt
\def\pmb#1{\setbox0=\hbox{#1}  \kern-.025em\copy0\kern-\wd0
 \kern.05em\copy0\kern-\wd0
  \kern-0.025em\raise.0433em\box0 }
\def\half{{\scriptstyle {1 \over 2}}}
\def\third{{\scriptstyle {1 \over 3}}}
\def\quarter{{\scriptstyle {1 \over 4}}}
\def\o{{\pmb o}}
\def\i{{\pmb{$\iota$}}}
\let\gt=\mapsto
\let\la=\lambda
\def\a{\overline a}
\def\daa{\nabla_{AA'}}
\def\ob{{\overline\o}}
\def\ib{{\overline\i}}
\def\T{{\bf T}}
\def\bp{{\bf p}}
\def\bq{{\bf q}}
\def\0{\pmb 0}
\def\1{\pmb 1}
\def\2{\pmb 2}
\def\3{\pmb 3}
\def\bPhi{{\pmb{$\Phi$}}}
\def\bPsi{{\pmb{$\Psi$}}}
\def\>{\phantom{A}}
\def\sym{{\sum_{sym}}}
\def\thorn{\hbox{\rm I}\kern-0.32em\raise0.35ex\hbox{\it o}}
\def\edth{\hbox{$\partial$\kern-0.25em\raise0.6ex\hbox{\rm\char'40}}}
\def\edthbar{\overline{\edth}}
\def\nedth{{\pmb\edth}}
\def\nedthp{{\pmb\edth'}}
\def\thornp{\thorn'}
\def\nthorn{{\pmb\thorn}}
\def\nthornp{{\pmb\thorn'}}
\def\edthp{\edth'}
\def\a{\alpha}
\def\b{\beta}
\def\g{\gamma}
\def\d{\delta}
\def\cd{{\cal D}}
\def\boeta{{\pmb{$\eta$}}}
\def\blambda{{\pmb{$\lambda$}}}
\def\etad{\boeta_{C_1\dots C_N C'_1\dots C'_{N'}}}
\def\p{{\bf p}}
\def\q{{\bf q}}
\def\K{{\bf K}}
\def\R{{\bf R}}
\def\S{{\bf S}}
\def\T{{\bf T}}
\def\I{{\bf I}}
\def\la{\lambda}
\def\lab{\overline\lambda}
\title{Obtaining a class of Type O pure radiation metrics with a cosmological constant, using invariant
operators.}
\author{S. Brian Edgar 
\\ Department of Mathematics,
 \\ Link\"{o}pings universitet\\ Link\"{o}ping,\\
Sweden S-581 83\\
 email: bredg@mai.liu.se
\and M.P. Machado Ramos
\\ Departamento de Matem\' atica 
\\ para a Ci\^encia e Tecnologia,
\\ Azur\'em 4810 Guimar$\tilde{\hbox {a}}$es,
\\
Universidade do Minho, 
\\ Portugal\\
 email: mpr@mct.uninho.pt}

\maketitle

\section*{Abstract.} 

Using the generalised invariant formalism we derive a class of conformally flat spacetimes whose Ricci tensor has a pure radiation and a Ricci scalar component.  The method used is a development of the methods used earlier for pure radiation spacetimes of Petrov types O and N respectively. In this paper we demonstrate how to handle, in the generalised invariant formalism,  spacetimes with isotropy freedom and rich Killing vector structure.  Once the spacetimes have been constructed, it is straightforward to deduce their Karlhede classification: the Karlhede algorithm terminates at the fourth derivative order, and the spacetimes all have one degree of null isotropy and three, four or five Killing vectors.

\

{\bf PACS numbers:} \ 0420,\ 1127

\

\section{Introduction}

\subsection{Conformally flat pure radiation spacetimes}

As pointed out in \cite{barnes} there are a number of interesting aspects to the complete class of conformally flat pure radiation spacetimes which have been presented in \cite{edlud1}, \cite{edludL}. Firstly, these spacetimes are of interest in their own right, having explicit physical interpretation, which has been investigated in  \cite{gr1}.  Secondly, the complete class of these spacetimes was found by a new  integration procedure  \cite{edlud1} within the GHP formalism \cite{ghp};  this GHP approach supplied an alternative, and in some senses, simpler approach  compared to the more complicated NP methods \cite{np} of finding exact solutions which required keeping track of  a lot of gauge and coordinate transformations. 
(A previous investigation of this class of spaces  using the NP formalism had overlooked this general case, and only identified  a subclass of these spaces (the Wils spacetime \cite{wils})).

Thirdly these spacetimes have provided interesting laboratories to test computer algebra programmes, such as those used in implementing the Karlhede algorithm \cite{karl}, \cite{kamac} for classifying spacetimes.  The Wils spacetime \cite{wils} was the first spacetime whose Karlhede algorithm required the determination of the Riemann tensor's fourth covariant derivative \cite{kou}, and although the complete class of all conformally flat pure radiation spacetimes did not require higher than fourth order derivatives for its classification, Skea \cite{skea1}  has emphasised the non-trivial didactic value of this  complete class. The   classification by Skea  \cite{skea1} via the Karlhede algorithm of the complete class revealed a mistake 
in the CLASSI, \cite{classi} programme when dealing with a rather subtle aspect of the freedom of a one-parameter group of null rotations; in addition, 
the   classification  of the complete class  in \cite{skea1} provided  a finer subdivision using discrete information than had been exploited before,  and sugggested the possibility of a refinement of the Karlhede classification algorithm, in general. 
Moreover,  when  this class of spacetimes was also used to demonstrate the GRtensor \cite{GR} implementation of the Karlhede algorithm, the result was in error \cite{pod}; this was because that  programme also failed to successfully interpret  a one-parameter group of null rotations.

Another interesting aspect of these spacetimes is that they provided the first demonstration of an integration method \cite{edvic} involving the generalised invariant formalism (GIF) of Machado Ramos and Vickers  \cite{maria1}, \cite{maria2}, \cite{maria3}. Furthermore, it was demonstrated that, having  generated the spacetimes in GIF, it was quite straightforward to deduce directly, by hand, their Karlhede algorithm from the existing calculations \cite{edvic}.

In \cite{skea1}, Skea argues, from theoretical reasons,  that conformally flat pure radiation fluids are among the most likely spacetimes to require high-order derivatives in their classification, and so it would appear natural to investigate some closely related classes in a search for other spacetimes requiring higher derivatives; adding a Ricci scalar term suggests itself as the most  obvious   generalisation.  In some classes of spacetimes the addition of a cosmological constant makes little significant difference, e.g. the Robinson-Trautman class, but in a variety of spacetimes investigated recently its introduction creates a significant difference.

Such a generalisation is appealing for  other reasons too:  such spaces will still have a physical interpretation within a theory which includes the cosmological constant, and they will complement those recent investigations of spacetimes of Petrov types D, II, III and N with a cosmological constant \cite{czmc}, 
\cite{carm1}, \cite{carm2}, \cite{ozs},  \cite{bic1}, \cite{bic2}, \cite{pod}, \cite{gr1}, \cite{gr2}; 
such spaces will also   provide  further, and possibly even stiffer, tests of the computer programmes for the Karlhede algorithm;  in addition, such spaces will provide 
opportunities for a deeper understanding of how to exploit the GIF formalism and its associated techniques, including its usefullness  for the Karlhede algorithm.

\subsection{GIF integration procedure}

The GIF integration procedure \cite{edvic}, \cite{edram} is a generalisation to the GIF \cite{maria1}, \cite{maria2}, \cite{maria3}  of an  integration method originally proposed by Held  \cite{held1}, \cite{held2}  and developed by Edgar and Ludwig \cite{edGHP}, \cite{edlud1}, \cite{edlud3}  in the GHP formalism \cite{ghp}.  It consists of manipulating all the  equations of the formalism in an attempt to construct a complete and involutive  set of tables  involving  first  derivative GIF spinor operators. The 'optimal  situation' to be sought is for this complete and involutive set to  include

$\bullet$  a  table for each of four real zero-weighted scalars, 

$\bullet$ a  table for one complex (non-trivially-)weighted scalar  

$\bullet$ a table for a second  spinor $\I_A$ (which is    not parallel to the first dyad spinor $\o_A$); such a spinor  should   emerge naturally from the calculations.  

An important element in this method is to recognise that much information resides in the GIF commutator equations (as well as in the GIF Ricci and Bianchi equations) and   in order that all this information is extracted it is essential that the commutators should be  applied explicitly to these five scalars, as well as to the new spinor ${\bf I}$ \cite{edvic}.

Of course, we can extract all the information by applying the commutators to
different (but essentially equivalent) combinations of these scalars and
spinor; however the particular choices above are best suited to our
integration procedure since the four $\{0,0\}$ weighted real scalars will
become the coordinates, the complex weighted scalar gives the spin and boost  gauge, while
the spinor $\mathbf{I}$ will be identified with the second dyad spinor ${%
\setbox0=\hbox{$\iota$} \kern-.025em\copy0\kern-\wd0 \kern.05em\copy0\kern-%
\wd0 \kern-0.025em\raise.0433em\box0 }$ in the GHP formalism.  Once these tables have been found, and the new spinor ${\bf I} $ identified with the second dyad spinor $\i$, the problem can be reduced to a purely scalar one in the GHP formaliism.

We emphasise that it is essential to have all of these scalars and the spinor ${\bf I}$, and to apply the commutators explicitly to all of them in turn, in order to be sure we have the complete information in the field equations.
In the 'optimal situation', all of these scalars and the spinor ${\bf I}$ will be {\it intrinsic} to GIF, and will be generated {\it directly} by manipulations and rearrangements in the GIF formalism; the generic class of conformally flat pure radiation spacetimes provided an example of this \cite{edvic}.  In less than optimal situations, some of the scalars and/or the spinor  ${\bf I}$ cannot be generated {\it directly} within   the GIF formalism; in such cases, it is essential that we create these 'missing' quantities, and so they   have to be introduced {\it indirectly, via their tables}, and since they are not intrinsic, we will refer to them as {\it complementary}..  The  special  non-generic subclass of the conformally flat pure radiation spacetimes provided an example of this \cite{edvic} since only three  intrinsic coordinate candidates were generated directly, and a fourth coordinate candidates had to be introduced indirectly via its table.

This technique of introducing a 'missing' coordinate candidate by its table had earlier been developed in the closely related integration procedure within the GHP formalism \cite{edlud1}, \cite{edlud3}, where it was understood that  the absence of such a coordinate candidate is associated with the presence of a Killing vector.  We often do not have to rely on guesswork to deduce such tables; in situations where the 'missing' coordinate candidates has a counterpart in the generic case, we can 'copy' the table structure of the intrinsic coordinate candidate  in the generic case,  but ensure that our new complementary coordinate candidate has no direct links with any other elements of the formalism.  On other occasions, we may not have the advantage of a generic case from which we can get hints, and in such cases we will need to carefully study the structure of the other equations, especially the commutators, to guess, and then check to confirm the validity of, an appropriate table;  this was the approach in \cite{edram}.

In GIF we can also encounter the situation where the second spinor  ${\bf I}$ fails to be generated directly and uniquely  in terms of intrinsic elements of GIF; this will happen in spacetimes which have one or two degrees of null isotropy freedom.

We wish to obtain more experience in  GIF of 'copying' tables in spacetimes rich in Killing vector structure, as well as in spacetimes with isotropy, and the spacetimes we will now investigate provide us with these possiblities.

So, in this paper we investigate a special class of conformally flat spaces whose  Ricci tensor has a pure radiation component as well as a Ricci scalar; equivalently these can be considered as conformally flat  pure radiation spacetimes with a cosmological constant.  Specifically we concentrate on a particular subclass  which has some interesting properties, and whose derivation and classification will require additional techniques and provide additional insights compared to the analysis of conformally flat  pure radiation spacetimes without the cosmological constant in \cite{edvic};   in particular, unlike  the latter, we will find that the  spacetimes identified in this paper have the isotropy freedom  of a one-parameter group of null rotations, as well as a richer Killing vector structure.

\section{GIF}\label{GIF}

In this section we will give 
 summaries from \cite{maria2} of the relevant parts of the GIF  which
are needed in this paper. The philosophy and general techniques of the GIF operator integration procedure have
been described in \cite{edvic}, \cite{edram} so we will not repeat these
discussions here, but rather we refer the reader to these references. 

In the GIF the role of the spin coefficients $\kappa$, $\sigma$, $\rho$ and $%
\tau$ is taken up by spinor quantities $\mathbf{K}$, $\mathbf{S}$, $\mathbf{R%
}$ and $\mathbf{T}$ given by
\begin{eqnarray}
\mathbf{K}&=&\kappa  \nonumber \\
\mathbf{S}_{A^\prime}&=&\sigma {\overline{\setbox0=\hbox{o} \kern-.025em\copy0%
\kern-\wd0 \kern.05em\copy0\kern-\wd0 \kern-0.025em\raise.0433em\box0 }}%
_{A^\prime}- \kappa{\overline{\setbox0=\hbox{$\iota$} \kern-.025em\copy0\kern-\wd%
0 \kern.05em\copy0\kern-\wd0 \kern-0.025em\raise.0433em\box0 }}_{A^\prime}
\nonumber \\
\mathbf{R}_A&=&\rho{\setbox0=\hbox{o} \kern-.025em\copy0\kern-\wd0 \kern.05em%
\copy0\kern-\wd0 \kern-0.025em\raise.0433em\box0 }_A-\kappa{\setbox0=%
\hbox{$\iota$} \kern-.025em\copy0\kern-\wd0 \kern.05em\copy0\kern-\wd0 \kern%
-0.025em\raise.0433em\box0 }_A  \nonumber \\
\mathbf{T}_{AA^\prime}&=& \tau {\setbox0=\hbox{o} \kern-.025em\copy0\kern-\wd%
0 \kern.05em\copy0\kern-\wd0 \kern-0.025em\raise.0433em\box0 }_A{\overline{\setbox%
0=\hbox{o} \kern-.025em\copy0\kern-\wd0 \kern.05em\copy0\kern-\wd0 \kern%
-0.025em\raise.0433em\box0 }}_{A^\prime}- \rho {\setbox0=\hbox{o} \kern%
-.025em\copy0\kern-\wd0 \kern.05em\copy0\kern-\wd0 \kern-0.025em\raise.0433em%
\box0 }_A{\overline{\setbox0=\hbox{$\iota$} \kern-.025em\copy0\kern-\wd0 \kern%
.05em\copy0\kern-\wd0 \kern-0.025em\raise.0433em\box0 }}_{A^\prime}- \sigma {%
\setbox0=\hbox{$\iota$} \kern-.025em\copy0\kern-\wd0 \kern.05em\copy0\kern-%
\wd0 \kern-0.025em\raise.0433em\box0 }_A{\overline{\setbox0=\hbox{o} \kern-.025em%
\copy0\kern-\wd0 \kern.05em\copy0\kern-\wd0 \kern-0.025em\raise.0433em\box0 }%
}_{A^\prime}+ \kappa{\setbox0=\hbox{$\iota$} \kern-.025em\copy0\kern-\wd0 %
\kern.05em\copy0\kern-\wd0 \kern-0.025em\raise.0433em\box0 }_A{\overline{\setbox0=%
\hbox{$\iota$} \kern-.025em\copy0\kern-\wd0 \kern.05em\copy0\kern-\wd0 \kern%
-0.025em\raise.0433em\box0 }}_{A^\prime}
\end{eqnarray}
Under a transformation of the spin frame given by
\begin{eqnarray}
{\setbox0=\hbox{o} \kern-.025em\copy0\kern-\wd0 \kern.05em\copy0\kern-\wd0 %
\kern-0.025em\raise.0433em\box0 }^A\gt\lambda{\setbox0=\hbox{o} \kern-.025em%
\copy0\kern-\wd0 \kern.05em\copy0\kern-\wd0 \kern-0.025em\raise.0433em\box0 }%
^A \qquad {\setbox0=\hbox{$\iota$} \kern-.025em\copy0\kern-\wd0 \kern.05em%
\copy0\kern-\wd0 \kern-0.025em\raise.0433em\box0 }^A\gt\lambda^{-1}{\setbox0=%
\hbox{$\iota$} \kern-.025em\copy0\kern-\wd0 \kern.05em\copy0\kern-\wd0 \kern%
-0.025em\raise.0433em\box0 }^A+\overline a {\setbox0=\hbox{o} \kern-.025em\copy0%
\kern-\wd0 \kern.05em\copy0\kern-\wd0 \kern-0.025em\raise.0433em\box0 }^A
\end{eqnarray}
these are therefore invariant under null rotations and have weight $\{\mathbf{%
p}, \mathbf{q}\}$ under spin and boost transformations given by
\begin{eqnarray}
\mathbf{K}&\gt& \lambda^3\overline\lambda \mathbf{K} \qquad : \quad\{\setbox0=\hbox{3} \kern-.025em\copy0\kern-\wd0 \kern.05em%
\copy0\kern-\wd0 \kern-0.025em\raise.0433em\box0 , \setbox0=\hbox{1} \kern%
-.025em\copy0\kern-\wd0 \kern.05em\copy0\kern-\wd0 \kern-0.025em\raise.0433em%
\box0 \}  \qquad \nonumber \\
\mathbf{S}_{A^{\prime}}&\gt& \lambda^3\mathbf{S}_{A^{\prime}} \qquad : \quad\{\setbox0=\hbox{3} \kern-.025em\copy0\kern-\wd0 \kern.05em%
\copy0\kern-\wd0 \kern-0.025em\raise.0433em\box0 , \setbox0=\hbox{0} \kern%
-.025em\copy0\kern-\wd0 \kern.05em\copy0\kern-\wd0 \kern-0.025em\raise.0433em%
\box0 \}  \nonumber \\
\mathbf{R}_A&\gt& \lambda^2\overline\lambda \mathbf{R}_A  \quad\ : \quad\{\setbox0=\hbox{2} \kern-.025em\copy0\kern-\wd0 \kern.05em%
\copy0\kern-\wd0 \kern-0.025em\raise.0433em\box0 , \setbox0=\hbox{1} \kern%
-.025em\copy0\kern-\wd0 \kern.05em\copy0\kern-\wd0 \kern-0.025em\raise.0433em%
\box0 \}\nonumber \\
\mathbf{T}_{AA^{\prime}}&\gt& \lambda^2 \mathbf{T}_{AA^{\prime}}\ \  \ : \quad\{\setbox0=\hbox{2} \kern-.025em\copy0\kern-\wd0 \kern.05em%
\copy0\kern-\wd0 \kern-0.025em\raise.0433em\box0 , \setbox0=\hbox{0} \kern%
-.025em\copy0\kern-\wd0 \kern.05em\copy0\kern-\wd0 \kern-0.025em\raise.0433em%
\box0 \}
\end{eqnarray}

The  GIF differential operators ${\setbox0=\hbox{\thorn} \kern-.025em%
\copy0\kern-\wd0 \kern.05em\copy0\kern-\wd0 \kern-0.025em\raise.0433em\box0 }
$, ${\setbox0=\hbox{\edth} \kern-.025em\copy0\kern-\wd0 \kern.05em\copy0\kern%
-\wd0 \kern-0.025em\raise.0433em\box0 }$, ${\setbox0=\hbox{\thorn} \kern%
-.025em\copy0\kern-\wd0 \kern.05em\copy0\kern-\wd0 \kern-0.025em\raise.0433em%
\box0 ^{\prime}}$ and ${\setbox0=\hbox{\edth} \kern-.025em\copy0\kern-\wd0 %
\kern.05em\copy0\kern-\wd0 \kern-0.025em\raise.0433em\box0 ^{\prime}}$, which
act on properly weighted symmetric spinors to produce symmetric spinors of
different valence and weight,  may all be defined in terms of
an auxiliary differential operator $\mathcal{D}_{ABA^{\prime}B^{\prime}}$
which is defined by
\begin{eqnarray}
&& \mathcal{D}_{ABA^{\prime}B^{\prime}}{\setbox0=\hbox{$\eta$} \kern-.025em%
\copy0\kern-\wd0 \kern.05em\copy0\kern-\wd0 \kern-0.025em\raise.0433em\box0 }%
_{C_1\dots C_N C^{\prime}_1\dots C^{\prime}_{N^{\prime}}}  \nonumber \\
& & = {\setbox0=\hbox{o} \kern-.025em\copy0\kern-\wd0 \kern.05em\copy0\kern-%
\wd0 \kern-0.025em\raise.0433em\box0 }_A{\overline{\setbox0=\hbox{o} \kern-.025em%
\copy0\kern-\wd0 \kern.05em\copy0\kern-\wd0 \kern-0.025em\raise.0433em\box0 }%
}_{A^{\prime}}\nabla_{BB^{\prime}}{\setbox0=\hbox{$\eta$} \kern-.025em\copy0%
\kern-\wd0 \kern.05em\copy0\kern-\wd0 \kern-0.025em\raise.0433em\box0 }%
_{C_1\dots C_N C^{\prime}_1\dots C^{\prime}_{N^{\prime}}}  \nonumber \\
& & \quad - (\mathbf{p}{\overline{\setbox0=\hbox{o} \kern-.025em\copy0\kern-\wd0 %
\kern.05em\copy0\kern-\wd0 \kern-0.025em\raise.0433em\box0 }}%
_{A^{\prime}}\nabla_{BB^{\prime}}{\setbox0=\hbox{o} \kern-.025em\copy0\kern-%
\wd0 \kern.05em\copy0\kern-\wd0 \kern-0.025em\raise.0433em\box0 }_A \ \ +%
\mathbf{q}{\setbox0=\hbox{o} \kern-.025em\copy0\kern-\wd0 \kern.05em\copy0%
\kern-\wd0 \kern-0.025em\raise.0433em\box0 }_A\nabla_{BB^{\prime}}{\overline{%
\setbox0=\hbox{o} \kern-.025em\copy0\kern-\wd0 \kern.05em\copy0\kern-\wd0 %
\kern-0.025em\raise.0433em\box0 }}_{A^{\prime}}){\setbox0=\hbox{$\eta$} \kern%
-.025em\copy0\kern-\wd0 \kern.05em\copy0\kern-\wd0 \kern-0.025em\raise.0433em%
\box0 }_{C_1\dots C_N C^{\prime}_1\dots C^{\prime}_{N^{\prime}}}  \label{ddi}
\end{eqnarray}
where ${\setbox0=\hbox{$\eta$} \kern-.025em\copy0\kern-\wd0 \kern.05em\copy0%
\kern-\wd0 \kern-0.025em\raise.0433em\box0 }$ has weight $\{\mathbf{p},%
\mathbf{q}\}$.

The GIF operators are obtained by contraction with ${\setbox0=\hbox{o} \kern%
-.025em\copy0\kern-\wd0 \kern.05em\copy0\kern-\wd0 \kern-0.025em\raise.0433em%
\box0 }$ and ${\overline{\setbox0=\hbox{o} \kern-.025em\copy0\kern-\wd0 \kern.05em%
\copy0\kern-\wd0 \kern-0.025em\raise.0433em\box0 }}$, and symmetrizing.
\begin{eqnarray}
({\setbox0=\hbox{\thorn} \kern-.025em\copy0\kern-\wd0 \kern.05em\copy0\kern-%
\wd0 \kern-0.025em\raise.0433em\box0 }{\setbox0=\hbox{$\eta$} \kern-.025em%
\copy0\kern-\wd0 \kern.05em\copy0\kern-\wd0 \kern-0.025em\raise.0433em\box0 }%
)_{AC_1\dots C_NA^{\prime}C^{\prime}_1\dots C^{\prime}_{N^{\prime}}} &=& {%
\sum_{sym}}{\setbox0=\hbox{o} \kern-.025em\copy0\kern-\wd0 \kern.05em\copy0%
\kern-\wd0 \kern-0.025em\raise.0433em\box0 }^B{\overline{\setbox0=\hbox{o} \kern%
-.025em\copy0\kern-\wd0 \kern.05em\copy0\kern-\wd0 \kern-0.025em\raise.0433em%
\box0 }}^{B^{\prime}}\mathcal{D}_{ABA^{\prime}B^{\prime}}{\setbox0=%
\hbox{$\eta$} \kern-.025em\copy0\kern-\wd0 \kern.05em\copy0\kern-\wd0 \kern%
-0.025em\raise.0433em\box0 }_{C_1\dots C_N C^{\prime}_1\dots
C^{\prime}_{N^{\prime}}} \\
({\setbox0=\hbox{\edth} \kern-.025em\copy0\kern-\wd0 \kern.05em\copy0\kern-%
\wd0 \kern-0.025em\raise.0433em\box0 }{\setbox0=\hbox{$\eta$} \kern-.025em%
\copy0\kern-\wd0 \kern.05em\copy0\kern-\wd0 \kern-0.025em\raise.0433em\box0 }%
)_{AC_1\dots C_NA^{\prime}B^{\prime}C^{\prime}_1\dots
C^{\prime}_{N^{\prime}}} &=& {\sum_{sym}}{\setbox0=\hbox{o} \kern-.025em\copy%
0\kern-\wd0 \kern.05em\copy0\kern-\wd0 \kern-0.025em\raise.0433em\box0 }^B%
\mathcal{D}_{ABA^{\prime}B^{\prime}}{\setbox0=\hbox{$\eta$} \kern-.025em\copy%
0\kern-\wd0 \kern.05em\copy0\kern-\wd0 \kern-0.025em\raise.0433em\box0 }%
_{C_1\dots C_N C^{\prime}_1\dots C^{\prime}_{N^{\prime}}} \\
({\setbox0=\hbox{\edth} \kern-.025em\copy0\kern-\wd0 \kern.05em\copy0\kern-%
\wd0 \kern-0.025em\raise.0433em\box0 ^{\prime}}{\setbox0=\hbox{$\eta$} \kern%
-.025em\copy0\kern-\wd0 \kern.05em\copy0\kern-\wd0 \kern-0.025em\raise.0433em%
\box0 })_{ABC_1\dots C_NA^{\prime}C^{\prime}_1\dots C^{\prime}_{N^{\prime}}}
&=& {\sum_{sym}}{\overline{\setbox0=\hbox{o} \kern-.025em\copy0\kern-\wd0 \kern%
.05em\copy0\kern-\wd0 \kern-0.025em\raise.0433em\box0 }}^{B^{\prime}}%
\mathcal{D}_{ABA^{\prime}B^{\prime}}{\setbox0=\hbox{$\eta$} \kern-.025em\copy%
0\kern-\wd0 \kern.05em\copy0\kern-\wd0 \kern-0.025em\raise.0433em\box0 }%
_{C_1\dots C_N C^{\prime}_1\dots C^{\prime}_{N^{\prime}}} \\
({\setbox0=\hbox{\thorn} \kern-.025em\copy0\kern-\wd0 \kern.05em\copy0\kern-%
\wd0 \kern-0.025em\raise.0433em\box0 ^{\prime}}{\setbox0=\hbox{$\eta$} \kern%
-.025em\copy0\kern-\wd0 \kern.05em\copy0\kern-\wd0 \kern-0.025em\raise.0433em%
\box0 })_{ABC_1\dots C_NA^{\prime}B^{\prime}C^{\prime}_1\dots
C^{\prime}_{N^{\prime}}} &=& {\sum_{sym}}\mathcal{D}_{ABA^{\prime}B^{\prime}}%
{\setbox0=\hbox{$\eta$} \kern-.025em\copy0\kern-\wd0 \kern.05em\copy0\kern-%
\wd0 \kern-0.025em\raise.0433em\box0 }_{C_1\dots C_N C^{\prime}_1\dots
C^{\prime}_{N^{\prime}}}\label{thornpdefn}
\end{eqnarray}
where $\displaystyle{\sum_{sym}}$ indicates symmetrization over all free
primed and unprimed indices. \medskip\noindent

In our calculations, we will need to know the result of contracting ${\setbox0=\hbox{\thorn} \kern%
-.025em\copy0\kern-\wd0 \kern.05em\copy0\kern-\wd0 \kern-0.025em\raise.0433em%
\box0 ^{\prime}}{\setbox0=\hbox{$\eta$} \kern-.025em\copy0\kern-\wd0 \kern%
.05em\copy0\kern-\wd0 \kern-0.025em\raise.0433em\box0 }$ with ${\setbox0=%
\hbox{o} \kern-.025em\copy0\kern-\wd0 \kern.05em\copy0\kern-\wd0 \kern%
-0.025em\raise.0433em\box0 }$ and ${\overline{\setbox0=\hbox{o} \kern-.025em\copy0%
\kern-\wd0 \kern.05em\copy0\kern-\wd0 \kern-0.025em\raise.0433em\box0 }}$ respectively, as well as analogous contractions on the other operators. In the case of a scalar field $\eta$, contracting (\ref{thornpdefn}) with ${\overline{\setbox0=\hbox{o} \kern-.025em\copy0%
\kern-\wd0 \kern.05em\copy0\kern-\wd0 \kern-0.025em\raise.0433em\box0 }}%
^{B^{\prime}}$ gives
\begin{eqnarray}
({\setbox0=\hbox{\thorn} \kern-.025em\copy0\kern-\wd0 \kern.05em\copy0\kern-%
\wd0 \kern-0.025em\raise.0433em\box0 ^{\prime}}\eta)_{ABA^{\prime}B^{\prime}}%
{\overline{\setbox0=\hbox{o} \kern-.025em\copy0\kern-\wd0 \kern.05em\copy0\kern-%
\wd0 \kern-0.025em\raise.0433em\box0 }}^{B^{\prime}} &= & {\scriptstyle {%
\frac{1 }{2}}}\{({\setbox0=\hbox{\edth} \kern-.025em\copy0\kern-\wd0 \kern%
.05em\copy0\kern-\wd0 \kern-0.025em\raise.0433em\box0 ^{\prime}}%
\eta)_{ABA^{\prime}}-q(\overline\tau{\setbox0=\hbox{o} \kern-.025em\copy0\kern-\wd%
0 \kern.05em\copy0\kern-\wd0 \kern-0.025em\raise.0433em\box0 }_A{\setbox0=%
\hbox{o} \kern-.025em\copy0\kern-\wd0 \kern.05em\copy0\kern-\wd0 \kern%
-0.025em\raise.0433em\box0 }_B{\overline{\setbox0=\hbox{o} \kern-.025em\copy0\kern%
-\wd0 \kern.05em\copy0\kern-\wd0 \kern-0.025em\raise.0433em\box0 }}%
_{A^{\prime}} -\overline\rho{\setbox0=\hbox{o} \kern-.025em\copy0\kern-\wd0 \kern%
.05em\copy0\kern-\wd0 \kern-0.025em\raise.0433em\box0 }_{(A}{\setbox0=%
\hbox{$\iota$} \kern-.025em\copy0\kern-\wd0 \kern.05em\copy0\kern-\wd0 \kern%
-0.025em\raise.0433em\box0 }_{B)}{\setbox0=\hbox{o} \kern-.025em\copy0\kern-%
\wd0 \kern.05em\copy0\kern-\wd0 \kern-0.025em\raise.0433em\box0 }%
_{A^{\prime}}  \nonumber \\
& & \quad -\overline\sigma{\setbox0=\hbox{o} \kern-.025em\copy0\kern-\wd0 \kern%
.05em\copy0\kern-\wd0 \kern-0.025em\raise.0433em\box0 }_A{\setbox0=\hbox{o} %
\kern-.025em\copy0\kern-\wd0 \kern.05em\copy0\kern-\wd0 \kern-0.025em\raise%
.0433em\box0 }_B{\overline{\setbox0=\hbox{$\iota$} \kern-.025em\copy0\kern-\wd0 %
\kern.05em\copy0\kern-\wd0 \kern-0.025em\raise.0433em\box0 }}_{A^{\prime}}
+\overline\kappa{\setbox0=\hbox{o} \kern-.025em\copy0\kern-\wd0 \kern.05em\copy0%
\kern-\wd0 \kern-0.025em\raise.0433em\box0 }_{(A}{\setbox0=\hbox{$\iota$} %
\kern-.025em\copy0\kern-\wd0 \kern.05em\copy0\kern-\wd0 \kern-0.025em\raise%
.0433em\box0 }_{B)}{\setbox0=\hbox{$\iota$} \kern-.025em\copy0\kern-\wd0 %
\kern.05em\copy0\kern-\wd0 \kern-0.025em\raise.0433em\box0 }%
_{A^{\prime}})\eta\}  \nonumber \\
& =& {\scriptstyle {\frac{1 }{2}}}\{({\setbox0=\hbox{\edth} \kern-.025em\copy%
0\kern-\wd0 \kern.05em\copy0\kern-\wd0 \kern-0.025em\raise.0433em\box0
^{\prime}}\eta)_{ABA^{\prime}}-q\overline{\mathbf{T}}_{A^{\prime}(A}{\setbox0=%
\hbox{o} \kern-.025em\copy0\kern-\wd0 \kern.05em\copy0\kern-\wd0 \kern%
-0.025em\raise.0433em\box0 }_{B)}\eta\}  \label{thornp.ob}
\end{eqnarray}
Although the definition of the differential operators appears quite complicated,
the fact that they take symmetric spinors to symmetric spinors means that
one can write down the equations in a more compact and index free notation.
In the compacted notation (\ref{thornp.ob}) becomes
\begin{eqnarray}
({\setbox0=\hbox{\thorn} \kern-.025em\copy0\kern-\wd0 \kern.05em\copy0\kern-%
\wd0 \kern-0.025em\raise.0433em\box0 ^{\prime}}\eta)\cdot{\overline{\setbox0=%
\hbox{o} \kern-.025em\copy0\kern-\wd0 \kern.05em\copy0\kern-\wd0 \kern%
-0.025em\raise.0433em\box0 }}={\scriptstyle {\frac{1 }{2}}}\{({\setbox0=%
\hbox{\edth} \kern-.025em\copy0\kern-\wd0 \kern.05em\copy0\kern-\wd0 \kern%
-0.025em\raise.0433em\box0 ^{\prime}}\eta)-q\overline\mathbf{T}\eta\}
\label{thornp.ob.com}
\end{eqnarray}
Similar calculations give
\begin{eqnarray}
({\setbox0=\hbox{\thorn} \kern-.025em\copy0\kern-\wd0 \kern.05em\copy0\kern-%
\wd0 \kern-0.025em\raise.0433em\box0 ^{\prime}}\eta)\cdot{\setbox0=\hbox{o} %
\kern-.025em\copy0\kern-\wd0 \kern.05em\copy0\kern-\wd0 \kern-0.025em\raise%
.0433em\box0 }={\scriptstyle {\frac{1 }{2}}}\{({\setbox0=\hbox{\edth} \kern%
-.025em\copy0\kern-\wd0 \kern.05em\copy0\kern-\wd0 \kern-0.025em\raise.0433em%
\box0 }\eta)-p\mathbf{T}\eta\}  \label{thornp.o}
\end{eqnarray}
\begin{eqnarray}
({\setbox0=\hbox{\edth} \kern-.025em\copy0\kern-\wd0 \kern.05em\copy0\kern-%
\wd0 \kern-0.025em\raise.0433em\box0 ^{\prime}}\eta)\cdot{\setbox0=\hbox{o} %
\kern-.025em\copy0\kern-\wd0 \kern.05em\copy0\kern-\wd0 \kern-0.025em\raise%
.0433em\box0 }={\scriptstyle {\frac{1 }{2}}}\{({\setbox0=\hbox{\thorn} \kern%
-.025em\copy0\kern-\wd0 \kern.05em\copy0\kern-\wd0 \kern-0.025em\raise.0433em%
\box0 }\eta)-p\mathbf{R}\eta\}  \label{edthp.o}
\end{eqnarray}
\begin{eqnarray}
({\setbox0=\hbox{\edth} \kern-.025em\copy0\kern-\wd0 \kern.05em\copy0\kern-%
\wd0 \kern-0.025em\raise.0433em\box0 }\eta)\cdot{\overline{\setbox0=\hbox{o} \kern%
-.025em\copy0\kern-\wd0 \kern.05em\copy0\kern-\wd0 \kern-0.025em\raise.0433em%
\box0 }}={\scriptstyle {\frac{1 }{2}}}\{({\setbox0=\hbox{\thorn} \kern-.025em%
\copy0\kern-\wd0 \kern.05em\copy0\kern-\wd0 \kern-0.025em\raise.0433em\box0 }%
\eta)-q\overline\mathbf{R}\eta\}  \label{edth.ob}
\end{eqnarray}
\begin{eqnarray}
({\setbox0=\hbox{\thorn} \kern-.025em\copy0\kern-\wd0 \kern.05em\copy0\kern-%
\wd0 \kern-0.025em\raise.0433em\box0 ^{\prime}}\eta)\cdot{\setbox0=\hbox{o} %
\kern-.025em\copy0\kern-\wd0 \kern.05em\copy0\kern-\wd0 \kern-0.025em\raise%
.0433em\box0 }\cdot{\overline{\setbox0=\hbox{o} \kern-.025em\copy0\kern-\wd0 \kern%
.05em\copy0\kern-\wd0 \kern-0.025em\raise.0433em\box0 }}={\scriptstyle {%
\frac{1 }{4}}}\{({\setbox0=\hbox{\thorn} \kern-.025em\copy0\kern-\wd0 \kern%
.05em\copy0\kern-\wd0 \kern-0.025em\raise.0433em\box0 }\eta)-p\mathbf{R}%
\eta-q\overline\mathbf{R}\eta\}  \label{thornp.oob}
\end{eqnarray}
For a spinor ${\setbox0=\hbox{$\eta$} \kern-.025em\copy0\kern-\wd0 \kern.05em%
\copy0\kern-\wd0 \kern-0.025em\raise.0433em\box0 }$ the above contractions
become more complicated. For example for a valence (1,0)-spinor ${\setbox0=%
\hbox{$\eta$} \kern-.025em\copy0\kern-\wd0 \kern.05em\copy0\kern-\wd0 \kern%
-0.025em\raise.0433em\box0 }_A$ of weight $\{\mathbf{p},\mathbf{q}\}$ we get
\begin{eqnarray}
({\setbox0=\hbox{\thorn} \kern-.025em\copy0\kern-\wd0 \kern.05em\copy0\kern-%
\wd0 \kern-0.025em\raise.0433em\box0 ^{\prime}}{\setbox0=\hbox{$\eta$} \kern%
-.025em\copy0\kern-\wd0 \kern.05em\copy0\kern-\wd0 \kern-0.025em\raise.0433em%
\box0 })\cdot{\setbox0=\hbox{o} \kern-.025em\copy0\kern-\wd0 \kern.05em\copy0%
\kern-\wd0 \kern-0.025em\raise.0433em\box0 }={\scriptstyle {\frac{1 }{3}}}\{{%
\setbox0=\hbox{\thorn} \kern-.025em\copy0\kern-\wd0 \kern.05em\copy0\kern-\wd%
0 \kern-0.025em\raise.0433em\box0 ^{\prime}}({\setbox0=\hbox{$\eta$} \kern%
-.025em\copy0\kern-\wd0 \kern.05em\copy0\kern-\wd0 \kern-0.025em\raise.0433em%
\box0 }\cdot{\setbox0=\hbox{o} \kern-.025em\copy0\kern-\wd0 \kern.05em\copy0%
\kern-\wd0 \kern-0.025em\raise.0433em\box0 })+ ({\setbox0=\hbox{\edth} \kern%
-.025em\copy0\kern-\wd0 \kern.05em\copy0\kern-\wd0 \kern-0.025em\raise.0433em%
\box0 }{\setbox0=\hbox{$\eta$} \kern-.025em\copy0\kern-\wd0 \kern%
.05em\copy0\kern-\wd0 \kern-0.025em\raise.0433em\box0 })-(\mathbf{p}-{%
1})\mathbf{T}{\setbox0=\hbox{$\eta$} \kern-.025em\copy0\kern-\wd0 \kern.05em%
\copy0\kern-\wd0 \kern-0.025em\raise.0433em\box0 }\}  \label{thornpI.o}
\end{eqnarray}
and 
\begin{eqnarray}  
({\setbox0=\hbox{\thorn} \kern-.025em\copy0\kern-\wd0 \kern.05em\copy0\kern-%
\wd0 \kern-0.025em\raise.0433em\box0 ^{\prime}}{\setbox0=\hbox{$\eta$} \kern%
-.025em\copy0\kern-\wd0 \kern.05em\copy0\kern-\wd0 \kern-0.025em\raise.0433em%
\box0 })\cdot{\overline{\setbox0=\hbox{o} \kern-.025em\copy0\kern-\wd0 \kern%
.05em\copy0\kern-\wd0 \kern-0.025em\raise.0433em\box0 }}={\scriptstyle {\frac{1 }{3}}}\{{%
\setbox0=\hbox{\thorn} \kern-.025em\copy0\kern-\wd0 \kern.05em\copy0\kern-\wd%
0 \kern-0.025em\raise.0433em\box0 ^{\prime}}({\setbox0=\hbox{$\eta$} \kern%
-.025em\copy0\kern-\wd0 \kern.05em\copy0\kern-\wd0 \kern-0.025em\raise.0433em%
\box0 }\cdot\overline {\setbox0=\hbox{o} \kern-.025em\copy0\kern-\wd0 \kern.05em\copy0%
\kern-\wd0 \kern-0.025em\raise.0433em\box0 })+ ({\setbox0=\hbox{\edth} \kern%
-.025em\copy0\kern-\wd0 \kern.05em\copy0\kern-\wd0 \kern-0.025em\raise.0433em%
\box0 }^{\prime}{\setbox0=\hbox{$\eta$} \kern-.025em\copy0\kern-\wd0 \kern%
.05em\copy0\kern-\wd0 \kern-0.025em\raise.0433em\box0 })-\mathbf{q
}\mathbf{\overline T}{\setbox0=\hbox{$\eta$} \kern-.025em\copy0\kern-\wd0 \kern.05em%
\copy0\kern-\wd0 \kern-0.025em\raise.0433em\box0 }\}  \label{thornpI.obar}
\end{eqnarray}

An alternative way to define the GIF operators is via the GHP operators $\thorn, \edth,\edthp, \thornp$, and we  can write equation (\ref{ddi}) in the form
\begin{eqnarray}
& &\!\!\mathcal{D}_{ABA^{\prime}B^{\prime}}{\setbox0=\hbox{$\eta$} \kern%
-.025em\copy0\kern-\wd0 \kern.05em\copy0\kern-\wd0 \kern-0.025em\raise.0433em%
\box0 }_{C_1\dots C_N C^{\prime}_1\dots C^{\prime}_{N^{\prime}}}  \nonumber
\\
& &= (\hbox{\rm I}\kern-0.32em\raise0.35ex\hbox{\it o}^{\prime}{\setbox0=%
\hbox{$\eta$} \kern-.025em\copy0\kern-\wd0 \kern.05em\copy0\kern-\wd0 \kern%
-0.025em\raise.0433em\box0 }_{C_1\dots C_N C^{\prime}_1\dots
C^{\prime}_{N^{\prime}}}){\setbox0=\hbox{o} \kern-.025em\copy0\kern-\wd0 %
\kern.05em\copy0\kern-\wd0 \kern-0.025em\raise.0433em\box0 }_A{\setbox0=%
\hbox{o} \kern-.025em\copy0\kern-\wd0 \kern.05em\copy0\kern-\wd0 \kern%
-0.025em\raise.0433em\box0 }_B{\overline{\setbox0=\hbox{o} \kern-.025em\copy0\kern%
-\wd0 \kern.05em\copy0\kern-\wd0 \kern-0.025em\raise.0433em\box0 }}%
_{A^{\prime}}{\overline{\setbox0=\hbox{o} \kern-.025em\copy0\kern-\wd0 \kern.05em%
\copy0\kern-\wd0 \kern-0.025em\raise.0433em\box0 }}_{B^{\prime}}  \nonumber
\\
& &\ -(\hbox{$\partial$\kern-0.25em\raise0.6ex\hbox{\rm\char'40}}^{\prime}{%
\setbox0=\hbox{$\eta$} \kern-.025em\copy0\kern-\wd0 \kern.05em\copy0\kern-\wd%
0 \kern-0.025em\raise.0433em\box0 }_{C_1\dots C_N C^{\prime}_1\dots
C^{\prime}_{N^{\prime}}}){\setbox0=\hbox{o} \kern-.025em\copy0\kern-\wd0 %
\kern.05em\copy0\kern-\wd0 \kern-0.025em\raise.0433em\box0 }_A{\setbox0=%
\hbox{o} \kern-.025em\copy0\kern-\wd0 \kern.05em\copy0\kern-\wd0 \kern%
-0.025em\raise.0433em\box0 }_B{\overline{\setbox0=\hbox{o} \kern-.025em\copy0\kern%
-\wd0 \kern.05em\copy0\kern-\wd0 \kern-0.025em\raise.0433em\box0 }}%
_{A^{\prime}}{\overline{\setbox0=\hbox{$\iota$} \kern-.025em\copy0\kern-\wd0 \kern%
.05em\copy0\kern-\wd0 \kern-0.025em\raise.0433em\box0 }}_{B^{\prime}}-(%
\hbox{$\partial$\kern-0.25em\raise0.6ex\hbox{\rm\char'40}}{\setbox0=%
\hbox{$\eta$} \kern-.025em\copy0\kern-\wd0 \kern.05em\copy0\kern-\wd0 \kern%
-0.025em\raise.0433em\box0 }_{C_1\dots C_N C^{\prime}_1\dots
C^{\prime}_{N^{\prime}}}){\setbox0=\hbox{o} \kern-.025em\copy0\kern-\wd0 %
\kern.05em\copy0\kern-\wd0 \kern-0.025em\raise.0433em\box0 }_A{\setbox0=%
\hbox{$\iota$} \kern-.025em\copy0\kern-\wd0 \kern.05em\copy0\kern-\wd0 \kern%
-0.025em\raise.0433em\box0 }_B{\overline{\setbox0=\hbox{o} \kern-.025em\copy0\kern%
-\wd0 \kern.05em\copy0\kern-\wd0 \kern-0.025em\raise.0433em\box0 }}%
_{A^{\prime}} {\overline{\setbox0=\hbox{o} \kern-.025em\copy0\kern-\wd0 \kern.05em%
\copy0\kern-\wd0 \kern-0.025em\raise.0433em\box0 }}_{B^{\prime}}  \nonumber
\\
& &\ \ -(\hbox{\rm I}\kern-0.32em\raise0.35ex\hbox{\it o}{\setbox0=%
\hbox{$\eta$} \kern-.025em\copy0\kern-\wd0 \kern.05em\copy0\kern-\wd0 \kern%
-0.025em\raise.0433em\box0 }_{C_1\dots C_N C^{\prime}_1\dots
C^{\prime}_{N^{\prime}}}){\setbox0=\hbox{o} \kern-.025em\copy0\kern-\wd0 %
\kern.05em\copy0\kern-\wd0 \kern-0.025em\raise.0433em\box0 }_A{\setbox0=%
\hbox{$\iota$} \kern-.025em\copy0\kern-\wd0 \kern.05em\copy0\kern-\wd0 \kern%
-0.025em\raise.0433em\box0 }_B{\overline{\setbox0=\hbox{o} \kern-.025em\copy0\kern%
-\wd0 \kern.05em\copy0\kern-\wd0 \kern-0.025em\raise.0433em\box0 }}%
_{A^{\prime}}{\overline{\setbox0=\hbox{$\iota$} \kern-.025em\copy0\kern-\wd0 \kern%
.05em\copy0\kern-\wd0 \kern-0.025em\raise.0433em\box0 }}_{B^{\prime}}
\nonumber \\
& & \ \ \ \ +(\mathbf{p}{\setbox0=\hbox{$\iota$} \kern-.025em\copy0\kern-\wd%
0 \kern.05em\copy0\kern-\wd0 \kern-0.025em\raise.0433em\box0 }_A{\overline{\setbox%
0=\hbox{o} \kern-.025em\copy0\kern-\wd0 \kern.05em\copy0\kern-\wd0 \kern%
-0.025em\raise.0433em\box0 }}_{A^{\prime}}\mathbf{T}_{BB^{\prime}} +\mathbf{q%
}{\setbox0=\hbox{o} \kern-.025em\copy0\kern-\wd0 \kern.05em\copy0\kern-\wd0 %
\kern-0.025em\raise.0433em\box0 }_A{\overline{\setbox0=\hbox{$\iota$} \kern-.025em%
\copy0\kern-\wd0 \kern.05em\copy0\kern-\wd0 \kern-0.025em\raise.0433em\box0 }%
}_{B^{\prime}}\overline{\mathbf{T}}_{B^{\prime}B}){\setbox0=\hbox{$\eta$} \kern%
-.025em\copy0\kern-\wd0 \kern.05em\copy0\kern-\wd0 \kern-0.025em\raise.0433em%
\box0 }_{C_1\dots C_N C^{\prime}_1\dots C^{\prime}_{N^{\prime}}}
\end{eqnarray}
where $\hbox{\rm I}\kern-0.32em\raise0.35ex\hbox{\it o}^{\prime}$, $%
\hbox{$\partial$\kern-0.25em\raise0.6ex\hbox{\rm\char'40}}^{\prime}$, $%
\hbox{$\partial$\kern-0.25em\raise0.6ex\hbox{\rm\char'40}}$ and $\hbox{\rm I}%
\kern-0.32em\raise0.35ex\hbox{\it o}$ are the ordinary GHP operators applied
to spinors.

In the case of a scalar field this gives
\begin{eqnarray}
(\nthornp\eta)_{ABA'B'}&= &
(\thornp\eta)\o_A\o_B\ob_{A'}\ob_{B'}
-(\edthp\eta-q\bar\tau\eta)\o_A\o_B\ob_{(A'}\ib_{B')}
\nonumber \\
& & \   -(\edth\eta-p\tau\eta)\o_{(A}\i_{B)}\ob_{A'}\ob_{B'} +(\thorn\eta-p\rho\eta-q\bar\rho\eta)\o_{(A}\i_{B)}\ob_{(A'}\ib_{B')}
\nonumber \\
& & \   -p\sigma\i_A\i_B\ob_{A'}\ob_{B'}-q\bar\sigma\o_A\o_B\i_{A'}\i_{B'}\nonumber \\
& & \ +p\kappa\i_A\i_B\ob_{(A'}\ib_{B')}+q\bar\kappa\o_{(A}\i_{B)}\ib_{A'}\ib_{B'} \label{thornp}
\end{eqnarray}
\begin{eqnarray}
(\nedthp\eta)_{ABA'}&=&
(\edthp\eta)\o_A\o_B\ob_{A'}
-(\thorn\eta-p\rho\eta)\o_{(A}\i_{B)}\ob_{A'}
\nonumber \\
& & \   +q\bar\sigma\o_A\o_B\i_{A'}
-p\kappa\i_A\i_B\ob_{A'}-q\bar\kappa\o_{(A}\i_{B)}\ib_{A'} \label{edthp}
\end{eqnarray}
\begin{eqnarray}
(\nedth\eta)_{AA'B'}&=&
(\edth\eta)\o_A\ob_{A'}\ob_{B'}
-(\thorn\eta-q\bar\rho\eta)\o_{A}\ob_{(A'}\ib_{B')}
\nonumber \\
& & \   +p\sigma\i_A\ob_{A'}\ob_{B'}
-p\kappa\i_A\ob_{(A'}\ib{B')}-q\bar\kappa\o_{A}\ib_{A'}\ib_{B'}\label{edth}
\end{eqnarray}
\begin{eqnarray}
(\nthorn\eta)_{AA'}&=
(\thorn\eta)\o_A\o_B
+p\kappa\i_A\ob_{A'}-q\bar\kappa\o_{A}\ib_{A'}\label{thorn}\ .
\end{eqnarray}
These equations will enable us to transfer from GIF to GHP formalism.

\medskip\noindent

The Ricci equations, Bianchi equations and the commutators in the GIF are given in \cite{maria2}. This complete system of equations is completely
equivalent to Einstein's equations, and to find solutions to Einstein's
equations this system will therefore have to be completely integrated.
However, in view of the more complicated nature of the operators in this
formalism, some of the information which resided in the Ricci equations in
NP and/or GHP formalisms is contained implicitly within the commutators in this
formalism; in particular these commutators contain inhomogeneous terms
explicitly dependent on the weight and valence of the spinor on which they
act.

\section{The equations}\label{equations}

We are concerned with the Petrov type O pure radiation spaces with non-zero
Ricci scalar. In the usual way, we choose ${\setbox0=\hbox{o} \kern-.025em%
\copy0\kern-\wd0 \kern.05em\copy0\kern-\wd0 \kern-0.025em\raise.0433em\box0 }%
_A$ to be aligned with the propogation direction of the radiation, so that
the Ricci spinor takes the form
\begin{eqnarray}
{\setbox0=\hbox{$\Phi$} \kern-.025em\copy0\kern-\wd0 \kern.05em\copy0\kern-%
\wd0 \kern-0.025em\raise.0433em\box0 }_{ABA^{\prime}B^{\prime}}=\Phi{\setbox%
0=\hbox{o} \kern-.025em\copy0\kern-\wd0 \kern.05em\copy0\kern-\wd0 \kern%
-0.025em\raise.0433em\box0 }_A{\setbox0=\hbox{o} \kern-.025em\copy0\kern-\wd%
0 \kern.05em\copy0\kern-\wd0 \kern-0.025em\raise.0433em\box0 }_B{\overline{\setbox%
0=\hbox{o} \kern-.025em\copy0\kern-\wd0 \kern.05em\copy0\kern-\wd0 \kern%
-0.025em\raise.0433em\box0 }}_{A^{\prime}}{\overline{\setbox0=\hbox{o} \kern%
-.025em\copy0\kern-\wd0 \kern.05em\copy0\kern-\wd0 \kern-0.025em\raise.0433em%
\box0 }}_{B^{\prime}}  \label{Phi}
\end{eqnarray}
where $\Phi (=\Phi_{22})$ is a real scalar field of weight $\{2,2\}$; all
the other curvature components, except the Ricci scalar $\Lambda$, vanish.

For this class of spaces the well known property of the vanishing of the
spin coefficients $\kappa, \sigma, \rho$ means that in the GIF
\begin{eqnarray}
\mathbf{K} &=0  \nonumber  \label{K} \\
\mathbf{S} &=0  \nonumber  \label{S} \\
\mathbf{R} & =0  \label{R}
\end{eqnarray}
but
\begin{eqnarray}
\mathbf{T}_{AA^{\prime}}=\tau{\setbox0=\hbox{o} \kern-.025em\copy0\kern-\wd0 %
\kern.05em\copy0\kern-\wd0 \kern-0.025em\raise.0433em\box0 }_A{\overline{\setbox0=%
\hbox{o} \kern-.025em\copy0\kern-\wd0 \kern.05em\copy0\kern-\wd0 \kern%
-0.025em\raise.0433em\box0 }}_{A^{\prime}}
\end{eqnarray}

Notice that $\tau$ and $\Phi_{22}$ are both invariant under the group of
null rotations so that they can be used instead of their GIF spinor
equivalents; this gives a considerable simplification in the GIF notation.

The GIF equations are:

\medskip

(i) GIF Ricci equations:
\begin{eqnarray}  \label{ricci}
{\setbox0=\hbox{\thorn} \kern-.025em\copy0\kern-\wd0 \kern.05em\copy0\kern-%
\wd0 \kern-0.025em\raise.0433em\box0 } \tau &=& 0 \\
{\setbox0=\hbox{\edth} \kern-.025em\copy0\kern-\wd0 \kern.05em\copy0\kern-\wd%
0 \kern-0.025em\raise.0433em\box0 } \tau &=& \tau^2 \\
{{\setbox0=\hbox{\edth} \kern-.025em\copy0\kern-\wd0 \kern.05em\copy0\kern-%
\wd0 \kern-0.025em\raise.0433em\box0 }^\prime} \tau&=& \tau\overline{\tau}%
+2\Lambda
\end{eqnarray}

(ii) GIF Bianchi equations:
\begin{eqnarray}  \label{bianchi1}
{\setbox0=\hbox{\thorn} \kern-.025em\copy0\kern-\wd0 \kern.05em\copy0\kern-%
\wd0 \kern-0.025em\raise.0433em\box0 }\Phi &=& 0 \\
{\setbox0=\hbox{\edth} \kern-.025em\copy0\kern-\wd0 \kern.05em\copy0\kern-\wd%
0 \kern-0.025em\raise.0433em\box0 }\Phi &=& \tau\Phi \\
{{\setbox0=\hbox{\edth} \kern-.025em\copy0\kern-\wd0 \kern.05em\copy0\kern-%
\wd0 \kern-0.025em\raise.0433em\box0 }^\prime}\Phi &=& \overline{\tau}\Phi
\end{eqnarray}
\begin{eqnarray}  \label{bianchi2}
{\setbox0=\hbox{\thorn} \kern-.025em\copy0\kern-\wd0 \kern.05em\copy0\kern-%
\wd0 \kern-0.025em\raise.0433em\box0 }\Lambda &=& 0  \nonumber \\
{\setbox0=\hbox{\edth} \kern-.025em\copy0\kern-\wd0 \kern.05em\copy0\kern-\wd%
0 \kern-0.025em\raise.0433em\box0 }\Lambda &=& 0  \nonumber \\
{{\setbox0=\hbox{\edth} \kern-.025em\copy0\kern-\wd0 \kern.05em\copy0\kern-%
\wd0 \kern-0.025em\raise.0433em\box0 }^\prime}\Lambda &=& 0  \nonumber \\
{{\setbox0=\hbox{\thorn} \kern-.025em\copy0\kern-\wd0 \kern.05em\copy0\kern-%
\wd0 \kern-0.025em\raise.0433em\box0 }^\prime}\Lambda &=& 0
\end{eqnarray}

(iii) GIF commutators (applied to a general symmetric spinor {\boldmath$\eta$%
} of weight ${\bf \{p,q\}}$ and with $N$ unprimed and $N'$ primed indices):
\begin{eqnarray}  \label{comma}
({\setbox0=\hbox{\thorn} \kern-.025em\copy0\kern-\wd0 \kern.05em\copy0\kern-%
\wd0 \kern-0.025em\raise.0433em\box0 } {\setbox0=\hbox{\thorn} \kern-.025em%
\copy0\kern-\wd0 \kern.05em\copy0\kern-\wd0 \kern-0.025em\raise.0433em\box0 }%
^\prime - {\setbox0=\hbox{\thorn} \kern-.025em\copy0\kern-\wd0 \kern.05em%
\copy0\kern-\wd0 \kern-0.025em\raise.0433em\box0 }^\prime{\setbox0=%
\hbox{\thorn} \kern-.025em\copy0\kern-\wd0 \kern.05em\copy0\kern-\wd0 \kern%
-0.025em\raise.0433em\box0 })\mbox{\boldmath$\eta$} &=& (\overline{\tau}{%
\setbox0=\hbox{\edth} \kern-.025em\copy0\kern-\wd0 \kern.05em\copy0\kern-\wd%
0 \kern-0.025em\raise.0433em\box0 } + \tau{\setbox0=\hbox{\edth} \kern-.025em%
\copy0\kern-\wd0 \kern.05em\copy0\kern-\wd0 \kern-0.025em\raise.0433em\box0 }%
^\prime)\mbox{\boldmath$\eta$}+(\mathbf{p}-N)\Lambda \mbox{\boldmath$\eta$}+(%
\mathbf{q}-N^\prime)\Lambda \mbox{\boldmath$\eta$}   \\  \label{commb}
({\setbox0=\hbox{\thorn} \kern-.025em\copy0\kern-\wd0 \kern.05em\copy0\kern-%
\wd0 \kern-0.025em\raise.0433em\box0 }{\setbox0=\hbox{\edth} \kern-.025em%
\copy0\kern-\wd0 \kern.05em\copy0\kern-\wd0 \kern-0.025em\raise.0433em\box0 }
- {\setbox0=\hbox{\edth} \kern-.025em\copy0\kern-\wd0 \kern.05em\copy0\kern-%
\wd0 \kern-0.025em\raise.0433em\box0 }{\setbox0=\hbox{\thorn} \kern-.025em%
\copy0\kern-\wd0 \kern.05em\copy0\kern-\wd0 \kern-0.025em\raise.0433em\box0 }%
)\mbox{\boldmath$\eta$}&=& 2\Lambda(\mbox{\boldmath$\eta$}\cdot{\setbox0=%
\hbox{o} \kern-.025em\copy0\kern-\wd0 \kern.05em\copy0\kern-\wd0 \kern%
-0.025em\raise.0433em\box0 })  \\ \label{commc}
({\setbox0=\hbox{\thorn} \kern-.025em\copy0\kern-\wd0 \kern.05em\copy0\kern-%
\wd0 \kern-0.025em\raise.0433em\box0 }{\setbox0=\hbox{\edth} \kern-.025em%
\copy0\kern-\wd0 \kern.05em\copy0\kern-\wd0 \kern-0.025em\raise.0433em\box0 }^{\prime}
- {\setbox0=\hbox{\edth} \kern-.025em\copy0\kern-\wd0 \kern.05em\copy0\kern-%
\wd0 \kern-0.025em\raise.0433em\box0 }^{\prime}{\setbox0=\hbox{\thorn} \kern-.025em%
\copy0\kern-\wd0 \kern.05em\copy0\kern-\wd0 \kern-0.025em\raise.0433em\box0 }%
)\mbox{\boldmath$\eta$}&=& 2\Lambda(\mbox{\boldmath$\eta$}\cdot \overline{\o})  \\ 
 \label{commd}
({\setbox0=\hbox{\edth} \kern-.025em\copy0\kern-\wd0 \kern.05em\copy0\kern-%
\wd0 \kern-0.025em\raise.0433em\box0 }{\setbox0=\hbox{\edth} \kern-.025em%
\copy0\kern-\wd0 \kern.05em\copy0\kern-\wd0 \kern-0.025em\raise.0433em\box0 }%
^\prime - {\setbox0=\hbox{\edth} \kern-.025em\copy0\kern-\wd0 \kern.05em\copy%
0\kern-\wd0 \kern-0.025em\raise.0433em\box0 }^\prime{\setbox0=\hbox{\edth} %
\kern-.025em\copy0\kern-\wd0 \kern.05em\copy0\kern-\wd0 \kern-0.025em\raise%
.0433em\box0 })\mbox{\boldmath$\eta$} &=& -(\mathbf{p}-N)\Lambda %
\mbox{\boldmath$\eta$}+(\mathbf{q}-N^\prime)\Lambda \mbox{\boldmath$\eta$}
 \\  \label{comme}
({\setbox0=\hbox{\thorn} \kern-.025em\copy0\kern-\wd0 \kern.05em\copy0\kern-%
\wd0 \kern-0.025em\raise.0433em\box0 }^\prime{\setbox0=\hbox{\edth} \kern%
-.025em\copy0\kern-\wd0 \kern.05em\copy0\kern-\wd0 \kern-0.025em\raise.0433em%
\box0 } - {\setbox0=\hbox{\edth} \kern-.025em\copy0\kern-\wd0 \kern.05em\copy%
0\kern-\wd0 \kern-0.025em\raise.0433em\box0 }{\setbox0=\hbox{\thorn} \kern%
-.025em\copy0\kern-\wd0 \kern.05em\copy0\kern-\wd0 \kern-0.025em\raise.0433em%
\box0 }^\prime)\mbox{\boldmath$\eta$} &=& -\tau{\setbox0=\hbox{\thorn} \kern%
-.025em\copy0\kern-\wd0 \kern.05em\copy0\kern-\wd0 \kern-0.025em\raise.0433em%
\box0 }^\prime\mbox{\boldmath$\eta$} -\Phi(\mbox{\boldmath$\eta$}\cdot {%
\setbox0=\hbox{o} \kern-.025em\copy0\kern-\wd0 \kern.05em\copy0\kern-\wd0 %
\kern-0.025em\raise.0433em\box0 })
 \\  \label{commf}
({\setbox0=\hbox{\thorn} \kern-.025em\copy0\kern-\wd0 \kern.05em\copy0\kern-%
\wd0 \kern-0.025em\raise.0433em\box0 }^\prime{\setbox0=\hbox{\edth} \kern%
-.025em\copy0\kern-\wd0 \kern.05em\copy0\kern-\wd0 \kern-0.025em\raise.0433em%
\box0 }^{\prime} - {\setbox0=\hbox{\edth} \kern-.025em\copy0\kern-\wd0 \kern.05em\copy%
0\kern-\wd0 \kern-0.025em\raise.0433em\box0 }^{\prime}{\setbox0=\hbox{\thorn} \kern%
-.025em\copy0\kern-\wd0 \kern.05em\copy0\kern-\wd0 \kern-0.025em\raise.0433em%
\box0 }^\prime)\mbox{\boldmath$\eta$} &=& -\overline{\tau}{\setbox0=\hbox{\thorn} \kern%
-.025em\copy0\kern-\wd0 \kern.05em\copy0\kern-\wd0 \kern-0.025em\raise.0433em%
\box0 }^\prime\mbox{\boldmath$\eta$} -\Phi(\mbox{\boldmath$\eta$}\cdot \overline{\o })
\end{eqnarray}
where $(\mbox{\boldmath$\eta$} \cdot {\bf o})$ is the $(N-1,N')$-spinor $\mbox{\boldmath$\eta$}_{A_1....A_NA_1....A_{N'}}{\bf o}^{A_N}$ , and $(\mbox{\boldmath$\eta$} \cdot {\bf \bar o})$ is the $(N,N'-1)$-spinor $\mbox{\boldmath$\eta$}_{A_1....A_NA_1....A_{N'}}\bar {\bf o}^{A_{N'}}$, and if the contraction is not possible then these terms are set to zero

These GIF equations contain all the information for the type O pure
radiation metrics with non-zero Ricci scalar. We emphasize that we assume
throughout that constant $\Lambda\neq 0$ as well as $\tau\neq 0$.

In this paper we shall only be concerned with the special subclass where 
\begin{equation}\tau\bar\tau + \Lambda =0\label{subdefn}
\end{equation}
which of course means that we will only be considering {\it a negative cosmological constant}, and it will be convenient to write $\lambda\equiv \pm\sqrt{-\Lambda}$.

\subsection{Preliminary rearrangement.}\label{intne1}

The Riemann tensor and the spin coefficients supply three real scalars which
can easily be rearranged to give one real zero-weighted $(\tau\overline{\tau}%
)$ and two real weighted scalars, $\Phi$ and $\arg(\tau/\overline{\tau})$.
In this special case, the real zero-weighted scalar $(\tau\overline{\tau}
)= \lambda^2$ and is constant;
and  in order for convenient  presentation we use the weighted scalars
\begin{equation}
{P}=\sqrt{\frac{\tau}{\overline{\tau}} },  \label{P}
\end{equation}
\begin{equation}
{Q}={\sqrt{\Phi}}  \label{Q}
\end{equation}
where ${P}$ is a complex scalar of weight $\{1,-1\}$ and ${P}\overline {P} =1$;
 ${Q}$ is a real scalar of weight $\{-1,-1\}$. (As well as $\Phi = {{%
{Q}^2}}\ne 0 \ne \Lambda$, we are assuming $\tau={\lambda {P}}\ne 0 $,
and so each of ${P},\ {Q}$, will always be defined and different from zero.)

These particular choices enable us to replace the Ricci and Bianchi equations with the
one equation
\begin{eqnarray}  \label{partialtablePbarQ}
{\setbox0=\hbox{\thorn} \kern-.025em\copy0\kern-\wd0 \kern.05em\copy0\kern-%
\wd0 \kern-0.025em\raise.0433em\box0 } (\overline {P}{Q}) &=& 0  \nonumber \\
{\setbox0=\hbox{\edth} \kern-.025em\copy0\kern-\wd0 \kern.05em\copy0\kern-\wd%
0 \kern-0.025em\raise.0433em\box0 } (\overline {P}{Q})&=&-\lambda Q /2
\nonumber \\
{{\setbox0=\hbox{\edth} \kern-.025em\copy0\kern-\wd0 \kern.05em\copy0\kern-%
\wd0 \kern-0.025em\raise.0433em\box0 }^\prime} (\overline {P}{Q}) &=&3\lambda {Q}
\overline{{P}}^2/2  \label{partialtablePQ}
\end{eqnarray}
bearing in mind that $\lambda$ is constant.

These spacetimes are clearly a very good example of a situation where very little explicit information is given via the Ricci and Bianchi equations; but, on the otherhand, we will find that a lot of additional information is  given implicitly via the commutators (\ref{comma}) -- (\ref{commf}).

\section{The integration procedure:  
 the generic case.}\label{intne}

\subsection{Constructing a table for ${\bf I}$  and applying commutators to ${\bf I}$.}\label{intne2}

For our integration procedure we begin by completing the partial table (\ref{partialtablePQ}) for the $\{-2,0\}$ weighted scalar $\overline{P}{Q}$,
\begin{eqnarray}  \label{tablePbarQ}
{\setbox0=\hbox{\thorn} \kern-.025em\copy0\kern-\wd0 \kern.05em\copy0\kern-%
\wd0 \kern-0.025em\raise.0433em\box0 } (\overline {P}{Q}) &=& 0  \nonumber \\
{\setbox0=\hbox{\edth} \kern-.025em\copy0\kern-\wd0 \kern.05em\copy0\kern-\wd%
0 \kern-0.025em\raise.0433em\box0 } (\overline {P}{Q})&=&-\lambda {Q}/2
\nonumber \\
{{\setbox0=\hbox{\edth} \kern-.025em\copy0\kern-\wd0 \kern.05em\copy0\kern-%
\wd0 \kern-0.025em\raise.0433em\box0 }^\prime} (\overline {P}{Q}) &=& 3\lambda {Q}%
\overline{{P}}^2/2  \nonumber \\
{{\setbox0=\hbox{\thorn} \kern-.025em\copy0\kern-\wd0 \kern.05em\copy0\kern-%
\wd0 \kern-0.025em\raise.0433em\box0 }^\prime} (\overline {P}{Q}) &=& \overline P Q\,\mathbf{J}   \label{tablePQ}
\end{eqnarray}
where we have completed the table with some spinor $\mathbf{J}$, which is as
yet undetermined; the additional factors are simply to shorten the subsequent presentation.

We know from (\ref{thornp.ob}) and (\ref{thornp.o}) that
\begin{equation}
{{\setbox0=\hbox{\thorn} \kern-.025em\copy0\kern-\wd0 \kern.05em\copy0\kern-%
\wd0 \kern-0.025em\raise.0433em\box0 }^\prime} (\overline {P}{Q})\cdot {\overline{\bf o}}=
{{\setbox0=\hbox{\edth} \kern-.025em\copy0\kern-\wd0 \kern.05em\copy0\kern-
\wd0 \kern-0.025em\raise.0433em\box0 }^\prime} (\overline {P}{Q}) \label{cdotobar}
\end{equation}
\begin{eqnarray}
{{\setbox0=\hbox{\thorn} \kern-.025em\copy0\kern-\wd0 \kern.05em\copy0\kern-%
\wd0 \kern-0.025em\raise.0433em\box0 }^\prime} (\overline {P}{Q})\cdot {\bf  o}=  
{\setbox0=\hbox{\edth} \kern-.025em\copy0\kern-\wd0 \kern.05em\copy0\kern-\wd%
0 \kern-0.025em\raise.0433em\box0 } (\overline {P}{Q}) + 2\tau \overline {P}{Q}=
{\setbox0=\hbox{\edth} \kern-.025em\copy0\kern-\wd0 \kern.05em\copy0\kern-\wd%
0 \kern-0.025em\raise.0433em\box0 } (\overline {P}{Q}) +2\lambda{Q}
\end{eqnarray}
Substituting (\ref{tablePQ}) we can then write
\begin{eqnarray}
{\bf J}=  -3\lambda (P{\bf I} +\overline P \overline {\bf I})/2\end{eqnarray}
where we have introduced the spinor ${\bf I}$ with the following simple properties
 \begin{equation}
\mathbf{I}\cdot {\overline{\setbox0=\hbox{o} \kern-.025em\copy0\kern-\wd0 \kern%
.05em\copy0\kern-\wd0 \kern-0.025em\raise.0433em\box0 }} =0  \label{Iobar}
\end{equation}
and
\begin{equation}
\mathbf{I}\cdot {\setbox0=\hbox{o} \kern-.025em\copy0\kern-\wd0 \kern.05em%
\copy0\kern-\wd0 \kern-0.025em\raise.0433em\box0 } =-1  \label{Io}
\end{equation}
Hence $\mathbf{I} $ is a $(1,0)$ valence spinor, and from
\begin{eqnarray}
{\bf J}
_{ABA^{\prime}B^{\prime}} =-\Bigl(3\lambda P/2\Bigr)
\mathbf{I}_{(A}{\setbox0=\hbox{o} \kern-.025em\copy0\kern-\wd0 \kern.05em%
\copy0\kern-\wd0 \kern-0.025em\raise.0433em\box0 }_{B)} {\overline{\setbox0=%
\hbox{o} \kern-.025em\copy0\kern-\wd0 \kern.05em\copy0\kern-\wd0 \kern%
-0.025em\raise.0433em\box0 }}_{A^{\prime}}{\overline{\setbox0=\hbox{o} \kern%
-.025em\copy0\kern-\wd0 \kern.05em\copy0\kern-\wd0 \kern-0.025em\raise.0433em%
\box0 }}_{B^{\prime}} -\Bigl(3\lambda\overline{{P}}/2\Bigr)\overline{\mathbf{I}}%
_{(A^{\prime}}{\overline{\setbox0=\hbox{o} \kern-.025em\copy0\kern-\wd0 \kern.05em%
\copy0\kern-\wd0 \kern-0.025em\raise.0433em\box0 }}_{B^{\prime})}{\setbox0=%
\hbox{o} \kern-.025em\copy0\kern-\wd0 \kern.05em\copy0\kern-\wd0 \kern%
-0.025em\raise.0433em\box0 }_A{\setbox0=\hbox{o} \kern-.025em\copy0\kern-\wd%
0 \kern.05em\copy0\kern-\wd0 \kern-0.025em\raise.0433em\box0 }_B
\end{eqnarray}
we conclude that its weight is $\{-\setbox0=\hbox{1} \kern-.025em\copy0\kern-%
\wd0 \kern.05em\copy0\kern-\wd0 \kern-0.025em\raise.0433em\box0 ,\setbox0=%
\hbox{0} \kern-.025em\copy0\kern-\wd0 \kern.05em\copy0\kern-\wd0 \kern%
-0.025em\raise.0433em\box0 \}$.

It is important to note two properties of the new spinor ${\bf I}$.  Firstly, ${\bf I}$ can never be zero, nor parallel to $\o$. Secondly,  it is emphasised that the spinor   ${\bf I}$, as defined above,  is {\it not} given uniquely in terms of the elements of the GIF formalism and so is not an intrinsic spinor;  ${\bf I}$ is only defined up to the freedom of a one dimensional null rotation 
\begin{equation}
{\bf I}\to {\bf I}+ i  \epsilon \bar P \,  \o \label{nullfreedom}
\end{equation}
where $\epsilon$ is an arbitrary  real zero-weighted scalar.

It will be useful  to have separate tables for $P$ and $Q$, 
\begin{eqnarray}  \label{tableP}
\bthorn {P} &=&0  \nonumber \\
\bedth {P} &=& \lambda {P}^2  \nonumber \\
\bedthp {P} &=& -\lambda   \nonumber \\
\bthornp {P} &=& 0
\end{eqnarray}
\begin{eqnarray}  \label{tableQ}
\bthorn {Q} &=& 0  \nonumber \\
\bedth {Q}&=&\lambda {Q}{P}/2  \nonumber \\
\bedthp {Q} &=&\lambda {Q}\overline{P}/2
\nonumber \\
\bthornp {Q} &=&- \frac{3{Q}{}\lambda}{2}(P\, \mathbf{%
I}+\overline{P}\, \overline{\mathbf{I}})
\end{eqnarray}

\smallskip

When we apply the commutators  (\ref{comma}) -- (\ref{commf}) to the table for $P$ they are identically satisfied, and when we apply them to the table for $Q$ we obtain  
\begin{eqnarray}
\bthorn(P\, {\bf I} + \overline P\, \overline{\bf I} ) &=&-2\lambda  \nonumber \\
\bedth(P\,{\bf I} + \overline P\, \overline {\bf I})  &=&\lambda P(P{\bf I}+\overline P  \overline{\bf I}) \nonumber \\
\bedthp(P\, {\bf I} + \overline P\, \overline {\bf I})  &=&\lambda \overline P(P{\bf I}+\overline P  \overline{\bf I})
  \label{partialtableI+Ibar}
\end{eqnarray}
We can complete this table with
\begin{eqnarray}
\bthornp(P\, {\bf I} + \overline P\, \overline {\bf I})  &=&{\bf K}  \label{addpartialtableI+Ibar}
\end{eqnarray}
where the spinor ${\bf K}$ is  as yet undetermined. Following the same procedure as  for the table for $\overline P Q$, 
we find 

\begin{eqnarray}
{\bf K}= Q^2 K -\lambda (P{\bf I}+\overline{P}\, \overline{\bf I})^2
  \label{Kdefn}
\end{eqnarray}
where $K$ is a zero-weighted real scalar, as yet undetermined. 

Therefore we do not obtain {\it directly} a  table for ${\bf I}$ which we require in order to apply the commutators  to  ${\bf I}$; and there is clearly no other way that we can supplement this information on ${\bf I}$ {\it directly}, by manipulation or rearranging.  
(This is a different situation from the analysis in \cite{edvic} where by applying the commutators to the table for $\overline P Q$ we obtained a partial table --- for operators $\bthorn, \bedth, \bedthp$ --- directly.)

 However, it is essential that we do obtain a  table for a second spinor; therefore 
 we introduce one  particular spinor $ {\bf I \!\!\!\!\,* } $    from the class of spinors ${\bf I}$ (which we noted were defined up to the freedom (\ref{nullfreedom})), by its partial table
\begin{eqnarray}
\bthorn{\bf I \!\!\!\!\,* }   &=&-\lambda  \overline {P}\nonumber \\
{\setbox0=\hbox{\edth} \kern-.025em\copy0\kern-\wd0 \kern.05em\copy0\kern-\wd%
0 \kern-0.025em\raise.0433em\box0 } {\bf I \!\!\!\!\,* } &=& {\lambda \overline{{P}}}\,\overline{\bf I \!\!\!\!\,* }    \nonumber \\
{{\setbox0=\hbox{\edth} \kern-.025em\copy0\kern-\wd0 \kern.05em\copy0\kern-%
\wd0 \kern-0.025em\raise.0433em\box0 }^\prime} {\bf I \!\!\!\!\,* }  &=&{\lambda \overline{{P}}}\,{\bf I \!\!\!\!\,* }   \label{partialtableI}
\end{eqnarray}
This table is clearly consistent with (\ref{partialtableI+Ibar}); moreover, we can confirm that it satisfies the relevant commutators (\ref{commb}) -- (\ref{commd}).  

We can seek to complete this table with 
\begin{equation} \bthornp {\bf I \!\!\!\!\,* }  = {\bf W}
\end{equation}  
where the spinor ${\bf W}$ is  as yet undetermined.

Following the same procedure as  for the table for $\overline P Q$, 
and again using (\ref{thornpI.obar}) and (\ref{thornpI.o}), we construct  the   completed table for ${\bf I \!\!\!\!\,* } $
\begin{eqnarray}
\bthorn{\bf I \!\!\!\!\,* }   &=&-\lambda  \overline {P}\nonumber \\
{\setbox0=\hbox{\edth} \kern-.025em\copy0\kern-\wd0 \kern.05em\copy0\kern-\wd%
0 \kern-0.025em\raise.0433em\box0 } {\bf I \!\!\!\!\,* } &=& {\lambda \overline{{P}}}\,\overline{\bf I \!\!\!\!\,* }    \nonumber \\
{{\setbox0=\hbox{\edth} \kern-.025em\copy0\kern-\wd0 \kern.05em\copy0\kern-%
\wd0 \kern-0.025em\raise.0433em\box0 }^\prime} {\bf I \!\!\!\!\,* }  &=&{\lambda \overline{{P}}}\,{\bf I \!\!\!\!\,* } \nonumber
\\
{{\setbox0=\hbox{\thorn} \kern-.025em\copy0\kern-\wd0 \kern.05em\copy0\kern-%
\wd0 \kern-0.025em\raise.0433em\box0 }^\prime} {\bf I \!\!\!\!\,* }  &=&{{\overline {P}
{Q}^2}}W-
{\lambda P}
{\bf I \!\!\!\!\,* } ^2-\lambda \overline{P}{\bf I \!\!\!\!\,* } \,\overline{\bf I \!\!\!\!\,* }  \label{tabletildeI}
\end{eqnarray}
where $W$ is a zero-weighted {\it complex} scalar, as yet undetermined. This table is clearly also consistent with (\ref{addpartialtableI+Ibar}) and (\ref{Kdefn}) with $K=W+\overline W$.

The theory  requires that we apply the commutators to the table for $
{\bf I \!\!\!\!\,* }$, which yields a partial table for $W$,
\begin{eqnarray}
{\setbox0=\hbox{\thorn} \kern-.025em\copy0\kern-\wd0
\kern.05em\copy0\kern- \wd0 \kern-0.025em\raise.0433em\box0 } W
&=& 0  \nonumber \\
{\setbox0=\hbox{\edth} \kern-.025em\copy0\kern-\wd0
\kern.05em\copy0\kern-\wd 0 \kern-0.025em\raise.0433em\box0 } W
&=& -{P}\Bigl(1-\lambda(W+\bar W)\Bigr)
\nonumber
\\
{{\setbox0=\hbox{\edth} \kern-.025em\copy0\kern-\wd0
\kern.05em\copy0\kern- \wd0 \kern-0.025em\raise.0433em\box0
}^\prime} W &=& 0\label{tableW}
\end{eqnarray}
This partial table satisfies the relevant commutators (\ref{commb}), (\ref{commc}), (\ref{commd}), so therefore the table (\ref{tabletildeI}) for ${\bf I \!\!\!\!\,* } $ is completely compatible with the remainder of the equations, and so we can adopt ${\bf I \!\!\!\!\,* } $ as the second spinor. However, as emphasised earlier, this spinor ${\bf I \!\!\!\!\,* } $ is {\it not defined  uniquely in terms of elements of the GIF formalism}, and so is not intrinsic to the spacetime.

\medskip

So we have obtained two of the core elements required in our analysis --- a
weighted scalar $\overline{{P}}{Q}$ and a new spinor ${\bf I \!\!\!\!\,* }$ which is not
parallel to ${\setbox0=\hbox{o} \kern-.025em\copy0\kern-\wd0 \kern.05em\copy0%
\kern-\wd0 \kern-0.025em\raise.0433em\box0 }$ --- and constructed
their tables; in addition, we have   applied the commutators to these tables in order to extract additional
information which was implicit in the commutators .

\subsection{Completing all the tables and applying the commutators }

We also need tables for four zero-weighted real scalars. 
Putting
\begin{eqnarray}\label{Wdefn}
W=M-iB + 1/2\lambda
\end{eqnarray}
the partial table  for (complex) $W$ (\ref{tableW}) yields partial tables  for (real) $M$ and $B$ respectively; we complete in the usual way to get  
\begin{eqnarray}
\bthorn M &=& 0\nonumber \\
\bedth M &=& \lambda {P}M\nonumber \\
\bedthp M &=& \lambda \overline{P} M\nonumber \\
\bthornp M &=& {Q}M^{3/2}R-\lambda {P}M{\bf I \!\!\!\!\,* } -\lambda \overline{P} M \overline{ {\bf I \!\!\!\!\,* }}
\label{tableM}
\end{eqnarray}
\begin{eqnarray}
\bthorn B &=& 0\nonumber \\
\bedth B &=& i\lambda {P}M\nonumber \\
\bedthp B &=& -i\lambda \overline{P}M\nonumber \\
\bthornp B &=&  {Q}M^{1/2}(G+RB) -i\lambda {P}M{\bf I \!\!\!\!\,* }+i\lambda \overline{P}M\overline{\bf I \!\!\!\!\,* }\label{tableB}
\end{eqnarray}
where $R$ and $G$ as usual are real zero-weighted scalars, as yet undetermined. (We have chosen the particular form and factors on these terms to shorten subsequent presentation.) 

 When we apply the commutators to the above tables for $M$ and $B$ we
obtain the following partial tables for $R$ and $G$ respectively
\begin{eqnarray}
\bthorn R &=& 0\nonumber \\
\bedth R &=& 0 \nonumber \\
\bedthp R &=& 0\label{partialtableR}
\end{eqnarray}
\begin{eqnarray}
\bthorn G &=& 0\nonumber \\
\bedth G &=& 0 \nonumber \\
\bedthp G &=& 0 \ . \label{partialtableG}
\end{eqnarray}
Since the \ $\bthorn$ \  component is zero in all four tables, it is clear that these four scalars, $M,B, R, G$ are not functionally independent; however the possibility of the three scalars $M, B, R$ being functionally independent is not obviously ruled out.  Therefore tentatively adopting $M,B,R$ as our three coordinate candidates, we complete the table for $R$ in the usual way with the zero-weighted scalar ${Y}$, as yet undetermined, 
\begin{eqnarray}\label{tableR}
\bthorn R &=& 0\nonumber \\
\bedth R &=& 0 \nonumber \\
\bedthp R &=& 0 \nonumber \\
\bthornp R &= & {Q} {Y}M^{1/2} \end{eqnarray}
Application of the commutators to $R$ gives 
\begin{eqnarray}\label{partialtable{Y}}
\bthorn {Y} &=& 0\nonumber \\
\bedth {Y} &=& 0   .\nonumber \\
\bedthp {Y} &=& 0 \end{eqnarray}

\smallskip
It is clear that we have extracted all the information which is available {\it directly} from the tables for $P, Q$ and ${\bf I \!\!\!\!\,* } $; we have applied the commutators a number of times ending up with identical partial tables for $R, G, Y$ which means that they are functionally dependent on each other, and hence
no further amount of rearranging nor manipulation with the commutators on the tables will yield a fourth scalar functionally independent of the three coordinate candidates $M, B, R$.  Clearly we need a table with a non-zero \ $\bthorn$ \ component. (In \cite{edvic} we were able to get a hint as to the structure of the 'missing'  table from a comparison with the generic case, when a fourth table was generated directly; no such comparison is  available in this paper.)

So we will try and introduce a  real zero-weighted scalar $ N$ via a table which is consistent with the commutators.   Beginning by checking the simplest possibilities we are led to 
\begin{eqnarray}
\bthorn  N &=& \frac{M^{3/2}}{{Q}}\nonumber\\
\bedth  N&=& -\frac{M^{3/2}}{{Q}}{\bf {\overline{\bf I \!\!\!\!\,* }}}\nonumber\\
\bedthp  N &=&  - \frac{M^{3/2}}{Q} {\bf I \!\!\!\!\,* } 
 \label{partialtableN}
\end{eqnarray}
which we can confirm satisfies the relevant commutators (\ref{commb}), (\ref{commc}), (\ref{commd}); so we complete the table as  
\begin{eqnarray}
\bthorn  N &=& \frac{M^{3/2}}{{Q}}\nonumber\\
\bedth  N&=& -\frac{M^{3/2}}{{Q}}{\bf {\overline{\bf I \!\!\!\!\,* }}}\nonumber\\
\bedthp  N &=&  - \frac{M^{3/2}}{{Q}}{\bf I \!\!\!\!\,* }  \nonumber\\
 \bthornp  N&=&  \frac{Q M^{1/2}}{2} U+\frac{M^{3/2}}{{Q}}{\bf I \!\!\!\!\,* } \, {\bf {\overline{\bf I \!\!\!\!\,* }}}
 \label{tableN}
\end{eqnarray}
where $U$ is a real zero-weighted scalar, as yet undetermined.

Applying the remaining commutators gives the partial table for $U$,
\begin{eqnarray}
\bthorn U &=& \frac{3M^{3/2}}{Q}R\nonumber\\
\bedth U &=& -2PM(M+iB+1/2\lambda)- \frac{3M^{3/2}}{Q}R\,{\bf {\overline{\bf I \!\!\!\!\,* }}}\nonumber\\
\bedthp U &=& -2\overline PM(M-iB+1/2\lambda)-
 \frac{3M^{3/2}}{Q}R\,{\bf{\bf I \!\!\!\!\,* }} \
 \label{partialtableU}
\end{eqnarray}
When the relevant commutators (\ref{commb}), (\ref{commc}), (\ref{commd}) are applied to $U$ it is found that they are identically satisfied.  Therefore the introduction of $ N$ via its table (\ref{tableN}) is completely compatible with the remaining equations
and so $ N$ can be taken as our fourth coordinate candidate.

In summary, we note that we have now applied the commutators to the four zero-weighted scalars $M,B,R,N$ in addition to the weighted scalar $PQ$ and the spinor ${\bf I \!\!\!\!\,* } $. Therefore we have obtained all the information about this class of spacetimes in the form of explicit equations.  Clearly our tables for the   four zero-weighted scalars $M,B,R,N$, the weighted scalar $PQ$ and the spinor ${\bf I \!\!\!\!\,* }$,  are not complete and involutive {\it by themselves} since they also contain the additional  zero-weighted scalars $G,Y,U$. However, the requirement of  applying the commutators to  the four coordinate candidates ensured that we also have the  constraint equations given by the partial tables (\ref{partialtableG}),  (\ref{partialtable{Y}}),  (\ref{partialtableU}) for those additional scalars, which taken together with the tables  (\ref{tableM}),    (\ref{tableB}), (\ref{tableR}), (\ref{tableN}), (\ref{tableP}), (\ref{tableQ}), (\ref{tabletildeI}), supply a complete and involutive ssytem.

\subsection{ The tables in GHP scalar operators}

If we identify the spinor ${\bf I \!\!\!\!\,* } $ with the second dyad spinor $\i$ of the GHP formalism, then the above tables for $M$, $B$, $R$ and $ N$, together with the additional constraint  equations, (\ref{partialtableG}),  (\ref{partialtable{Y}}),  (\ref{partialtableU})  can all be translated into the GHP formalism with the usual GHP scalar operators using (\ref{thornp}) -- (\ref{thorn}):
\begin{eqnarray}
\thorn M &=& 0\nonumber \\
\edth M &=&\lambda {P}M\nonumber \\
\edthp M &=&\lambda\overline {P} M\nonumber \\
\thornp M &=& {Q}RM^{3/2} \label{GHPtableM}
\end{eqnarray}
\begin{eqnarray}
\thorn B &=& 0\nonumber \\
\edth B &=& i\lambda {P}M\nonumber \\
\edthp B &=& -i   \lambda\overline {P}M\nonumber \\
\thornp B &=& {Q}(G+RB) M^{1/2}  \label{GHPtableB}
\end{eqnarray}
\begin{eqnarray}
\thorn R &=& 0\nonumber \\
\edth R &=& 0 \nonumber \\
\edthp R &=& 0   \nonumber \\
\thornp R &=& {Q}{Y}M^{1/2} \label{GHPtableR}
\end{eqnarray}
\begin{eqnarray}
\thorn N &=& \frac{M^{3/2}}{{Q}}\nonumber\\
\edth  N&=& 0\nonumber\\
\edthp  N &=& 0\nonumber\\
 \thornp  N&=&  \frac{{Q}M^{1/2}}{2}U
 \label{GHPtableN}
\end{eqnarray}
where
\begin{eqnarray}
\thorn G &=& 0\nonumber \\
\edth G &=&0\nonumber \\
\edthp G &=&0
\label{GHPpartialtableG}
\end{eqnarray}
\begin{eqnarray}
\thorn {Y} &=& 0\nonumber \\
\edth {Y} &=& 0\nonumber \\
\edthp {Y} &=& 0
\label{GHPpartialtable{Y}}
\end{eqnarray}
\begin{eqnarray}
\thorn U &=&  \frac{3M^{3/2}}{Q}R\nonumber\\
\edth U &=& -2PM(M+iB+1/2\lambda)\nonumber\\
\edthp U &=& -2\overline PM(M-iB+1/2\lambda)\ .
 \label{GHPpartialtableU}
\end{eqnarray}

For completeness we add the separate GHP tables for $P$ and $Q$, 
\begin{eqnarray}  \label{GHPtableP}
\thorn {P} &=&0  \nonumber \\
\edth {P} &=& \lambda {P}^2  \nonumber \\
\edthp {P} &=& -\lambda   \nonumber \\
\thornp {P} &=& 0
\end{eqnarray}
\begin{eqnarray}  \label{GHPtableQ}
\thorn {Q} &=& 0  \nonumber \\
\edth {Q}&=&\lambda {Q}{P}/2  \nonumber \\
\edthp {Q} &=&\lambda {Q}\overline{P}/2
\nonumber \\
\thornp {Q} &=& 0
\end{eqnarray}

Before we can adopt the
coordinate candidates as coordinates, we must confirm that they
are functionally independent.  Assuming that $M\ne 0$, we can easily confirm that $B, R, N$ cannot be constant; an  examination of the determinant formed from the four
tables (\ref{GHPtableM}), (\ref{GHPtableB}),  (\ref{GHPtableR}) and (\ref{GHPtableN})   shows that the
four coordinate candidates are indeed functionally independent --- providing ${M} \ne 0 \ne Y$.
In the next subsection we will consider the case  $ M\ne 0 \ne Y$, while in the subsection following that we will consider the case  $ M\ne 0 = Y$ (and more generally $ M\ne 0 $, all $Y$).
The case $M=0$ (and more generally, all $M$, all $Y$) will be looked at separately in the next section.

\subsection{Using coordinate candidates as coordinates and constructing a metric}

Assuming for  this subsection that  ${M}\ne 0\ne Y$, we  make the obvious choice of the coordinate candidates as  the coordinates\[
r=  R,\qquad  n=  N, \qquad m=M, \qquad b=  B
\label{coords1}
\]
Note that the coordinates $r,m,b$ have direct and unique identification with their respective coordinate candidates $R,M,B$, whereas $n$  is defined by $N$ only up to an additive constant, because $N$ is introduced indirectly via its table.

We can now write down the tetrad vectors in these coordinates by means
of the tables (\ref{GHPtableM}), (\ref{GHPtableB}), (\ref{GHPtableR}) and (\ref{GHPtableN}),
\begin{eqnarray}
l^i &=& \frac{1}{{Q}} (0, \ m^{3/2}, \
0, \ 0)   \nonumber \\
m^i &=& {P}\Bigl(0, \ 0, \ \lambda{m}, \
i\lambda{m}\Bigr)  \nonumber \\
\overline {m}^i &=& \overline {P}\Bigl(0, \ 0, \ \lambda{m}, \
-i\lambda{m}\Bigr) \nonumber \\
n^i &=& {Q}\Bigl({Y}m^{1/2}, \
Um^{1/2}/2,\   r m^{3/2} ,\    \bigl(rb+G\bigr)m^{1/2}\Bigr)\label{frameA11}
\end{eqnarray}
where the equations (\ref{GHPpartialtableG}), (\ref{GHPpartialtable{Y}}) and
(\ref{GHPpartialtableU}) give the constraints for $ G$, ${Y}$ and
$U$ respectively in the chosen coordinate system.

It is clear from (\ref{GHPpartialtableG}) and (\ref{GHPpartialtable{Y}}) that the functions $G$ and  ${Y}$ are independent of all coordinates except $r$, and hence we have ${Y}=\nu_1(r)$ and ${G}=\nu_2(r)$ where $\nu_2(r)$ is a completely arbitrary function of $r$, whereas $\nu_1(r)$  is an arbitrary function of $r$ excluding the zero function.  

From   (\ref{GHPpartialtableU}), in this coordinate system,  we get the following differential equations
\begin{eqnarray}
U_{,n}&=& 3 r \nonumber \\
\lambda U_{,m}+i\lambda U_{,b}&=&-2(m+ib+1/2\lambda)
\end{eqnarray}
from which we find 
\begin{equation}
U  ={3 }r  n - \frac{  b^2}{\lambda}-
\frac{m^2}{\lambda}-\frac{m}{\lambda^2} +\nu_3(r)
 \label{Usim1}
\end{equation}
where $\nu_3(r)$ is a completely arbitrary function of $r$. 

It follows immediately from the equation
$$g^{ij}=2l^{(i}n^{j)}-2m^{(i}\bar m^{j)}
$$
that the metric $g^{ij}$, in $r, n,m,b$ coordinates,  is  given by
\begin{equation}
g^{ij}=  {m^2}\left(
\begin{array}{lccr}
0 &\nu_1(r)& 0 & 0 
\cr \nu_1(r) &
U & mr&
\bigl( r b+\nu_2(r)\bigr)\cr 0 &  mr&-{2}\lambda^2 & 0 \cr 0 &
\bigl( rb+ \nu_2(r)\bigr)
 & 0 &-{2}\lambda^2
\end{array}
\right)  \label{metric1}
\end{equation}
where $U$ is given by (\ref{Usim1}).

\section{The integration procedure: extending the generic case}\label{extend}

\subsection{Preliminaries}

We ruled out  ${Y}=0$ in the previous subsection, since in that case, from (\ref{GHPtableR}), $R$ is a constant and therefore we
cannot take  $R$ as a coordinate. On the otherhand we still have the possibility of getting a fourth coordinate candidate from $G$ or $U$.  Once we make such a choice, then we could continue in a similar manner as in the last section, building our tables, and hence the tetrad, around the four coordinate candidates.

However, if \textit{neither} of the other functions $G, U$ is functionally
independent of the original three coordinates, then it will \textit{not} be
possible to find a replacement candidate \textit{directly}; we emphasise that in such circumstances no additional independent quantities can be generated by any direct manipulations of the tables and the commutators.  In
such a situation we still need a replacement coordinate candidate in order to extract
the remaining information from the commutators.  So,  rather than treating the
special case ${Y}= 0$ separately, we will extend the generic result to include the
special case as well.

We will now show that a replacement candidate for $R$ can be found, and that by defining  this {\it complementary coordinate candidate} indirectly via its table, we can obtain a metric which includes all possible values for $R$, including zero.

\subsection{Finding a complementary coordinate candidate to replace  $R$, and constructing a metric}

The results in the previous section apply; the only difference here is that we \textit{interpret}
them differently. When we are interpretating our tables and
choosing our explicit coordinate candidates we will now consider
only the three zero-weighted real scalars $M,B, N$
as coordinate candidates while the zero-weighted scalar $R$
is not now included as a coordinate candidate, which means that it is no longer prevented from acquiring a constant value, even zero. A related change is that
since $R$ is no longer a coordinate candidate, we  need only its  \textit{partial} table from (\ref{partialtableR}),
\begin{eqnarray}
\bthorn R &=& 0\nonumber \\
\bedth R &=& 0 \nonumber \\
\bedthp R &=& 0\label{partialtableR1}
\end{eqnarray}
So, clearly we do not have our full quota of \textit{four}
coordinate candidates, but we do not wish to use any of the
remaining quantities from the tables, since it would involve the
additional assumption of that quantity being non-constant.
 So we have to introduce a {\it complementary} zero-weighted scalar, functionally
independent of the first three coordinate candidates, whose table
is consistent with the commutators. In fact, we get a strong hint
from the previous subsection, and consider the possibility of the
existence of a real zero-weighted scalar $\tilde R $, which satisfies the
table
\begin{eqnarray}
\bthorn \tilde R &=& 0\nonumber \\
\bedth \tilde R &=& 0\nonumber \\
\bedthp \tilde R &=& 0\nonumber \\
\bthornp \tilde R &=& {Q}M^{1/2} \label{tabletildeR}
\end{eqnarray}
We have defined\footnote{For easy reference, in an extended case, we will label by $\tilde X$ a complementary coordinate candidate which replaces a coordinate candidate $X$ in a generic case; but we emphasise this is not to imply any direct link between the two quantities, it simply points us to the source of the hint which suggested the table for the complementary coordinate candidate.}
our new coordinate candidate $\tilde R$ by a table which is essentially the same structure as the table (\ref{tableR}) for the coordinate candidate $R$ which it replaces; but, unlike $R$ in the previous sections,  $\tilde R$ has no direct links to any other quantities in the equations.  (In fact (\ref{tableR}) has a slightly different structure than (\ref{tabletildeR}); however, if we had retained an arbitrary function $\tilde Y(\tilde R)$ in the table (\ref{tabletildeR}) analagous to (\ref{tableR}), the simple coordinate transformation $\tilde R \to \int \tilde Y(\tilde R)\, d\tilde R$ 
reduces it to unity.)

This table  (\ref{tabletildeR}) is easily seen to satisfy all the commutators (\ref{comma}) -- (\ref{commf}) and to be  compatible with the other tables. 

The GHP table for $\tilde R$ can be obtained
from (\ref{tabletildeR}) by substituting the GIF operators with the GHP scalar 
operators in the usual way,
\begin{eqnarray}
\thorn \tilde R &=& 0\nonumber \\
\edth \tilde R &=& 0\nonumber \\
\edthp \tilde R &=& 0\nonumber \\
\thornp \tilde R &=& {Q} M^{1/2}\label{caseA1tildeRghp}
\end{eqnarray}
The GHP tables (\ref{GHPtableM}), (\ref{GHPtableB}), (\ref{GHPtableN}), 
for the other three coordinate candidates $M,B,N$  respectively remain unchanged, as do the GHP partial tables (\ref{GHPpartialtableG}) and (\ref{GHPpartialtableU}) for $G$ and $U$ .

 It can now easily be seen from the determinant of the respective tables that  $\tilde R $, $M$, $ B$ and $ N$
are functionally independent and therefore can be chosen as
coordinate candidates. From (\ref{partialtableR1}), the GHP partial  table for $R$ becomes 
\begin{eqnarray}
\thorn R &=& 0\nonumber \\
\edth R &=& 0 \nonumber \\
\edthp R &=& 0\label{partialtableRghp}
\end{eqnarray}
which means that $R$ is a function of only the one coordinate candidate, $\tilde R$. Clearly from (\ref{GHPpartialtableG}), $G$ is also   a function of only the one coordinate candidate, $\tilde R$.

Hence, by replacing the coordinate $r$ and its table with the coordinate $\tilde r$ and its table,  we are now able to generalise the  metric form (\ref{metric1}) given in the previous subsection  in the coordinates 
\[
\tilde r=  \tilde R,\qquad  n=  N, \qquad m=M, \qquad b=  B
\label{coords2}
\]
 Note that the coordinates $m,b$ have direct and unique identification with their respective coordinate candidates $M,B$, whereas $n$ and $\tilde r$   are defined by $N$ and $\tilde R$ respectively only up to an additive constant, because $N$ and $\tilde R$ are introduced indirectly via their respective tables. 

The metric, in coordinates $\tilde r,  n, m, b,$  is given by
\begin{equation}
g^{ij}=  {m^2} \left(
\begin{array}{lccr}
0 &1& 0 & 0 \cr 1 &
U &m\nu_4(\tilde r)&
\bigl(b\nu_4(\tilde r)+\nu_2(\tilde r)\bigr)\cr 0 & m\nu_4(\tilde r)&-{2}\lambda^2 & 0 \cr 0 &
\bigl(b\nu_4(\tilde r)+\nu_2(\tilde r)\bigr)
 & 0 &-{2}\lambda^2
\end{array}
\right)  \label{metric2}
\end{equation}
where $U$ is obtained by solving (\ref{GHPpartialtableU}) to obtain  
\begin{equation}
U  ={3} n\nu_4(\tilde r) - \frac{ b^2}{\lambda}-
\frac{m^2}{\lambda}-\frac{m}{\lambda^2} +\nu_3(\tilde r)
 \label{Usim2}
\end{equation}
and $\nu_2(\tilde r) (=G(\tilde r))$, $\nu_3(\tilde r)$ and $\nu_4(\tilde r) (=R(\tilde r))$ are all completely arbitrary functions of $\tilde r$, (including the zero function).

When we compare the   metric (\ref{metric1})  with the
above metric (\ref{metric2}), we can
easily see that the former is a special case of the latter, since the latter also includes the special case corresponding to $\nu_4(\tilde r) $ being a constant function, which was excluded from the former; in the case when $\nu_4(\tilde r) $ is not a constant function, by taking $r=\nu_4(\tilde r)$ we  can retrieve the previous result (\ref{metric1}).

\section{The integration procedure: the special case $M= 0$ leading to the complete solution}\label{complete}

\subsection{Preliminaries}

When we substitute $M=0$ into the two tables  (\ref {tableM}) and  (\ref {tableB})  the first table collapses completely, and the second reduces the function $B$ to a constant.  Hence there only exists the tables for ${P},{Q}$ which are unchanged from (\ref{tableP}) and (\ref{tableQ}), as well as the table  (\ref{tabletildeI}) for 
 ${\bf I \!\!\!\!\,* } $ which is simplified by the substitution $W= -iB+1/2\lambda$  where  $B$  is a constant.
 
This is the only {\it direct} information that we obtain for this subclass, and no further information can be generated by applying the commutators to these tables; there is not even one  explicit zero-weighted quantity to be a coordinate candidate.

So we will need to introduce {\it four} complementary coordinate candidates indirectly via their respective tables, and to ensure that these tables are compatible with the commutators (\ref{comma}) -- (\ref{commf}) and the other three  tables (\ref{tableP}), (\ref{tableQ}) and  (\ref{tabletildeI}).  

In the previous section when we wished to fill in a missing subclass corresponding to a missing coordinate from the generic case, rather than treating the special case separately, we found it easier to extend the generic result to include the special case as well:
to obtain a replacement coordinate we  introduced indirectly a complementary coordinate candidate via the structure of the table for the corresponding coordinate candidate in the generic case, with the complementary coordinates independent of the other elements in the formalism. The solution we obtained extended the generic case to include the special missing subclass.

In this section we will follow the same principles: 
 we will  introduce indirectly, via their tables,  {\it four} complementary coordinates,  none of which occur directly in any other parts of the formalism. The solution then obtained will be a further extension of the extended version found in  section \ref{extend} (which was itself an extension of the generic case in section  \ref{intne}), since it will also include the subclass $M=0$; in fact it will be the complete solution to the class of spacetimes we have been investigating.

\subsection{Finding four complementary candidates, and constructing the complete metric}

Taking the hint from the respective table structures  in  section \ref{extend}, (\ref{tableM}),  (\ref{tableB}), (\ref{tableN}), (\ref{tabletildeR}), we introduce the four complementary candidates $\tilde M$, $\tilde B$, $\tilde R$, and $\tilde N$ via the four tables
\begin{eqnarray}
\bthorn \tilde M &=& 0\nonumber \\
\bedth \tilde M &=& \lambda {P}\tilde M\nonumber \\
\bedthp \tilde M &=& \lambda \overline{P} \tilde M\nonumber \\
\bthornp \tilde M &=& Q\tilde M^{3/2} R -\lambda {P}\tilde M{\bf I \!\!\!\!\,* } -\lambda \overline{P} \tilde M\overline {\bf I \!\!\!\!\,* } 
\label{tabletildeM0}
\end{eqnarray}
  \begin{eqnarray}
\bthorn \,\tilde{R}  &=& 0\nonumber \\
\bedth \,\tilde{ R}  &=& 0 \nonumber \\
\bedthp \,\tilde{R}  &=& 0\nonumber \\
\bthornp \,\tilde{ R}  &=& {Q}\tilde M^{1/2}
\label{tabletildeR0}
\end{eqnarray}
\begin{eqnarray}
\bthorn \tilde {B} &=& 0\nonumber \\
\bedth \tilde {B} &=& i\lambda {P}\tilde M\nonumber \\
\bedthp \tilde {B} &=& -i\lambda \overline{P}\tilde M\nonumber \\
\bthornp \tilde {B} &=& {Q}\tilde M^{1/2} R\tilde {B}-i\lambda {P}\tilde M{\bf I \!\!\!\!\,* } +i\lambda \overline{P}\tilde M\overline {\bf I \!\!\!\!\,* } 
\label{tabletildeB0}
\end{eqnarray}
\begin{eqnarray}
\bthorn \tilde N &=& \frac{\tilde M^{3/2}}{{Q}}\nonumber\\
\bedth \tilde N&=& -\frac{\tilde M^{3/2}}{{Q}}\overline{\bf I \!\!\!\!\,* } \nonumber\\
\bedthp \tilde N &=&  - \frac{\tilde M^{3/2}}{{Q}}{\bf I \!\!\!\!\,* } \nonumber\\
 \bthornp \tilde N&=& \frac{ {Q}\tilde M^{1/2}}{2} V +\frac{\tilde M^{3/2}}{{ Q}}\,{\bf I \!\!\!\!\,* }\, \overline{\bf I \!\!\!\!\,* } 
 \label{tabletildeN0}
\end{eqnarray}

 (Table (\ref{tableB}) has a slightly different structure than (\ref{tabletildeB0}); however, if we had retained an arbitrary function $ G(\,\tilde{ R})$ in the table (\ref{tabletildeB0}), the simple coordinate transformation $$\tilde B \to \tilde B + \exp(\tilde R^2/2) \int   G(\,\tilde{R})\, \exp(-\tilde R^2/2) \,d\tilde{ R}$$  reduces it to zero.)

Note that the scalar quantity $R$ has been left undetermined, rather than equating it to the coordinate candidate $\tilde R$ as was done in the generic case, since we are seeking to extend further the  generic result's extension, given in  the last section.

Alongside these are the original three tables (\ref{tableP}), (\ref{tableQ}) and
(\ref{tabletildeI}) (with $M$ and $B$ substituting for $W$ via (\ref{Wdefn})) for $P,Q, {\bf I \!\!\!\!\,* }$ respectively; it is emphasised that the scalar functions $M$ and $B$ in  the table (\ref{tabletildeI}) for ${\bf I \!\!\!\!\,* } $  have no direct link with the complementary coordinate candidates $\tilde M$, $\tilde B$. 

Applying the commutators to $P,Q$ gave the table (\ref{tabletildeI}) for ${\bf I \!\!\!\!\,* }$, and applying the commutators to   ${\bf I \!\!\!\!\,* }$ gives  the partial tables for the unknown functions,
\begin{eqnarray}
\bthorn  M &=& 0\nonumber \\
\bedth M &=& \lambda {P} M\nonumber \\
\bedthp M &=& \lambda \overline{P}  M\label{partialtableM0}
\end{eqnarray}
\begin{eqnarray}
\bthorn  {B} &=& 0\nonumber \\
\bedth {B} &=& i\lambda {P} M\nonumber \\
\bedthp  {B} &=& -i\lambda \overline{P} M\label{partialtableB0}
\end{eqnarray}

It remains to apply the commutators to the  four complementary coordinate candidates $\tilde M, \tilde B, \tilde R, \tilde N,$ defined by their tables above, and the only non-trivial results are,
\begin{eqnarray}
\bthorn  R &=& 0\nonumber \\
\bedth R &=& 0\nonumber \\
\bedthp R &=& 0\label{partialtableR0}
\end{eqnarray}
\begin{eqnarray}
\bthorn V &=& \frac{3\tilde M^{3/2}}{Q} R\nonumber\\
\bedth  V &=& -2P\tilde M(M+iB+1/2\lambda)- \frac{3\tilde M^{3/2}}{Q} R\,{\overline{\bf I \!\!\!\!\,* }} \nonumber\\
\bedthp  V &=& -2\overline P\tilde M(M-iB+1/2\lambda)-
 \frac{3\tilde M^{3/2}}{Q} R \, {\bf I \!\!\!\!\,* }  \
 \label{partialtabletildeV}
\end{eqnarray}

The relevant commutators (\ref{commb}) -- (\ref{commd}) are consistent when applied to the partial tables for  $R$ and $V$, and so our choices of the four complementary coordinate candidates $ \tilde M, \tilde R, \tilde B, \tilde N$ is permissable.

Once again, when all the tables and partial tables are considered, we have a complete and involutive set of equations.

We next translate all of these equations into their GHP versions in the usual way,
\begin{eqnarray}
\thorn \tilde M &=& 0\nonumber \\
\edth \tilde M &=& \lambda {P}\tilde M\nonumber \\
\edthp \tilde M &=& \lambda \overline{P} \tilde M\nonumber \\
\thornp \tilde M &=& Q\tilde M^{3/2}R\label{GHPtabletildeM0}
\end{eqnarray}
  \begin{eqnarray}
\thorn \,\tilde{ R}  &=& 0\nonumber \\
\edth \,\tilde{ R}  &=& 0 \nonumber \\
\edthp \,\tilde{R}  &=& 0\nonumber \\
\thornp \,\tilde{R}  &=& {Q}\tilde M^{1/2}
\label{GHPtabletildeR0}
\end{eqnarray}
\begin{eqnarray}
\thorn \tilde {B} &=& 0\nonumber \\
\edth \tilde {B} &=& i\lambda {P}\tilde M\nonumber \\
\edthp \tilde {B} &=& -i\lambda \overline{P}\tilde M\nonumber \\
\thornp \tilde {B} &=& {Q}\tilde M^{1/2}\tilde {B}R \label{GHPtabletildeB0}
\end{eqnarray}
\begin{eqnarray}
\thorn \tilde N &=& \frac{\tilde M^{3/2}}{{Q}}\nonumber\\
\edth \tilde N&=& 0 \nonumber\\
\edthp \tilde N &=&  0 \nonumber\\
 \thornp \tilde N&=& \frac{ {Q}\tilde M^{1/2}}{2} V  \label{GHPtabletildeN0}
\end{eqnarray}
alongside the GHP partial tables for  $R$,  $M$,  $B$  and $V$ respectively
\begin{eqnarray}
\thorn  R &=& 0\nonumber \\
\edth R &=& 0\nonumber \\
\edthp R &=& 0\label{GHPpartialtableR0}
\end{eqnarray}
\begin{eqnarray}
\thorn  M &=& 0\nonumber \\
\edth M &=& \lambda {P} M\nonumber \\
\edthp M &=& \lambda \overline{P}  M\label{GHPpartialtableM0}
\end{eqnarray}
\begin{eqnarray}
\thorn  {B} &=& 0\nonumber \\
\edth {B} &=& i\lambda {P} M\nonumber \\
\edthp  {B} &=& -i\lambda \overline{P} M\label{GHPpartialtableB0}
\end{eqnarray}
\begin{eqnarray}
\thorn V &=& \frac{3\tilde M^{3/2}}{Q} R\nonumber\\
\edth V &=& -2P\tilde M(M+iB+1/2\lambda)\nonumber\\
\edthp V &=& -2\overline P\tilde M(M-iB+1/2\lambda) \label{GHPpartialtabletildeU0}
\end{eqnarray}
Checking the determinant of the four tables (\ref{GHPtabletildeM0}), (\ref{GHPtabletildeR0}), (\ref{GHPtabletildeB0}), (\ref{GHPtabletildeN0}), confirms that these four scalars are functionally independent, and 
so we  make the obvious choice of the complementary coordinate candidates as  the coordinates\[
\tilde r=  \tilde{ R},\qquad \tilde n=  \tilde N, \qquad \tilde m= \tilde M, \qquad \tilde b=  \tilde B
\label{coords3}
\]
Note that all of the coordinates are defined  only up to an additive constant, because they have been  introduced indirectly via their tables. 

 We can now write down the tetrad vectors in these coordinates by means
of the tables (\ref{GHPtabletildeR0}), (\ref{GHPtabletildeM0}), (\ref{GHPtabletildeN0}) and (\ref{GHPtabletildeB0}),
\begin{eqnarray}
l^i &=& \frac{1}{{Q}} (0, \ \tilde m^{3/2}, \
0, \ 0)   \nonumber \\
m^i &=& {P}\Bigl(0, \ 0, \ \lambda{\tilde m}, \
i\lambda{\tilde m}\Bigr)  \nonumber \\
\overline {m}^i &=& \overline {P}\Bigl(0, \ 0, \ \lambda{\tilde m}, \
-i\lambda{\tilde m}\Bigr) \nonumber \\
n^i &=& {Q}\Bigl(\tilde m^{1/2}, \
V\tilde m^{1/2}/2,\   R \tilde m^{3/2} ,\    R\tilde b\tilde m^{1/2}\Bigr)\label{tetrad3}
\end{eqnarray}
From  (\ref{GHPpartialtableR0}) 
we know that $R$ is a function of $\tilde r$ only, and we will write  $R(\tilde r) = \nu_4(\tilde r) $ which is a completely arbitrary function of $\tilde r$.

Solving the equations (\ref{GHPpartialtableM0}), (\ref{GHPpartialtableB0}) and
(\ref{GHPpartialtabletildeU0}), respectively,  gives 
\begin{equation}
M=\tilde m \nu_5(\tilde r)
\end{equation}
\begin{equation}
B= \tilde b \nu_5(\tilde  r)+\nu_6(\tilde  r)
\end{equation}
\begin{equation}\label{Usim3}
V  ={3} \tilde n\nu_4(\tilde r) - \frac{ \nu_5(\tilde r)}{\lambda}(\tilde b^2+
\tilde m^2)-\frac{2\nu_6(\tilde r)\tilde b}{\lambda} -\frac{\tilde m}{\lambda^2} +\nu_3(\tilde r)
\end{equation}
where $\nu_3(\tilde r), \nu_5(\tilde r), \nu_6(\tilde r), $ are all also completely arbitrary functions of $\tilde r$. 

It follows immediately from the equation
$$g^{ij}=2l^{(i}n^{j)}-2m^{(i}\bar m^{j)}
$$
that the metric $g^{ij}$, in $\tilde r, \tilde n,\tilde m, \tilde b$ coordinates,  is  given by
\begin{equation}
g^{ij}=  {\tilde m^2}\left(
\begin{array}{lccr}
0 &1& 0 & 0 
\cr 1&
V & \tilde m\nu_4( \tilde r)&
\tilde b \nu_4 (\tilde r)
\cr 0 &  \tilde m\nu_4(\tilde r)&-{2}\lambda^2 & 0
 \cr 0 &
\tilde b  \nu_4( \tilde r)
 & 0 &-{2}\lambda^2
\end{array}
\right)  \label{metric3}
\end{equation}
where $V$ is given by (\ref{Usim3}).

This is the complete metric. The special case $M=0$  which was omitted in section \ref{extend} is given by $\nu_5(\tilde r)=0$, and the extended generic version (\ref{metric2})  from section \ref{extend} can be found when $\nu_5(\tilde r)\ne 0$ by a simple coordinate change.

\section{Karlhede classification}\label{Karlhede}

The efficiency of the GIF for investing the Karlhede classification \cite{karl}, \cite{kamac} of a metric has been discussed in \cite{edvic}; here we now apply to the class of spacetimes constructed in this paper
 the same procedure as was developed in \cite{edvic}. 
We consider  the complete solution given by (\ref{metric3}).

At zeroth order,
\begin{equation}
 \Phi = Q^2 \label{zero}
\end{equation}
At first order,
\begin{eqnarray}  \label{first}
\bthorn {\Phi} &=& 0  \nonumber \\
\bedth {\Phi}&=&\lambda {Q}^2{P}  \nonumber \\
\bedthp {\Phi} &=&\lambda {Q}^2\overline{P}
\nonumber \\
\bthornp {\Phi} &=&- 3{Q}^2\lambda( P\,\mathbf{
I} +\overline{P}\,\overline{\mathbf{I}})
\end{eqnarray}
We can solve for $Q$ at zeroth order and for $P$ and $( P\,\mathbf{
I} +\overline{P}\,\overline{\mathbf{I}})$ at first order; therefore ${\bf I}$ is not uniquely determined, and  it has  clearly the gauge freedom of a one parameter subgroup of null rotations.

At second order,  we find that the only non-zero expressions which give anything other than terms in $P$ and $Q$ are
\begin{eqnarray}  \label{second} 
\bthornp \bedth{\Phi} &=&- 3{Q}^4\lambda^2( P\,\mathbf{
I} +\overline{P}\,\overline{\mathbf{I}})
 \nonumber \\
\bthornp \bthornp {\Phi} &=&-3Q^4\lambda \Bigl(2\tilde m \nu_5(\tilde r) +1/\lambda\Bigr)  +12{Q}^2\lambda^2\bigr(P \,  {\bf I  } +  \overline P \, \overline{\bf I  }\bigl)^2
\end{eqnarray}
Providing $\nu_5(\tilde r) \ne 0$, we can  solve for a first essential coordinate $(\tilde m \nu_5(\tilde r))$, but  the gauge freedom of ${\bf I}$
remains unchanged.

At third order,  we have the equation
\begin{eqnarray}  \label{third1}
\bthornp \bthornp \bthornp {\Phi} &= -6Q^5\lambda  \tilde m^{3/2}\Bigl( \nu_5'(\tilde r)+\nu_5(\tilde r)\nu_4(\tilde r)\Bigr)  \nonumber\\ 
\qquad\quad  &+6 Q^4\lambda^2 \Bigr(15\tilde m \nu_5(\tilde r)+7/\lambda\Bigl) \bigr(P \,  {\bf I  } +  \overline P  \,\overline{\bf I }\bigl) \nonumber\\ &\quad -24{Q}^2\lambda^3\bigr(P  \, {\bf I } +  \overline P  \, \overline{\bf I  }\bigl)^3
\end{eqnarray}
where $\nu_5'(\tilde r)=\frac{d \nu_5(\tilde r)} {d \tilde r \  \  \  \   }$.   We can now solve for a second essential coordinate $\Bigl(\tilde m^{3/2}\bigl( \nu_5'(\tilde r)+\nu_5(\tilde r)\nu_4(\tilde r)\bigr)\Bigr) $, which is obviously functionally independent of the first, {\it in general}; clearly $\tilde m$ and $\tilde r$ are essential coordinates, {\it in general}.   However, for all third order values, the gauge freedom of ${\bf I }$
still remains unchanged.

At fourth order, we find, {\it in general},  that there are no new functionally independent scalars generated; moreover, the gauge freedom of a one parameter subgroup of null rotations for ${\bf I  }$ remains unchanged.  Hence the algorithm terminates at fourth order, {\it in general}, with two essential coordinates.

\medskip

However, there is a special case, at third order, since the   second proposed essential coordinate $\Bigl(\tilde m^{3/2}\bigl( \nu_5'(\tilde r)+\nu_5(\tilde r)\nu_4(\tilde r)\bigr)\Bigr) $ is functionally dependent on the first essential coordinate $(\tilde m \nu_5(\tilde r))$ in the case when $\bigl( \nu_5'(\tilde r)+\nu_5(\tilde r)\nu_4(\tilde r)\bigr)=k \bigl(\nu_5(\tilde r)\bigr)^{3/2} $, i.e.,
\begin{equation}      \nu_4(\tilde r)=\Bigl(k\bigl(\nu_5(\tilde r)\bigr)^{3/2}- \nu_5'(\tilde r)\Bigr)/ \nu_5(\tilde r)
\label{special}
\end{equation}
where $k$ is a constant.

All the other derivatives at third order fail to generate any new essential coordinate, and the gauge freedom of ${\bf I }$
still remains unchanged. Therefore, for this generic case, the algorithm terminates at third order.  

\medskip

Finally, we note that when $\nu_5(\tilde r) =0  $ there is no new information at second order, and so the algorithm terminates there.

\medskip

We can sum up as follows:

$\bullet$\  When $\nu_5(\tilde r)\ne 0$  and when $ \nu_4(\tilde r)\ne \Bigl(k\bigl(\nu_5(\tilde r)\bigr)^{3/2}- \nu_5'(\tilde r)\Bigr)/ \nu_5(\tilde r)
$ , we need to go to fourth order, and this subclass has two essential coordinates $\tilde m, \tilde r$, and one degree of isotropy and hence three Killing vectors.

$\bullet$ \ If $\nu_5(\tilde r)\ne 0$ and $ \nu_4(\tilde r)=\Bigl(k\bigl(\nu_5(\tilde r)\bigr)^{3/2}- \nu_5'(\tilde r)\Bigr)/ \nu_5(\tilde r)
$   no new information is given at third order, and the subclass has one essential coordinate $\tilde m \nu_5(\tilde r)$, and one degree of isotropy and hence four Killing vectors.

$\bullet$\  If $\nu_5(\tilde r) = 0$  no new information is given at second order, and the subclass has no essential coordinates, and one degree of isotropy and hence five Killing vectors.

\medskip

We note that there are no further subclasses depending on the values of the arbitrary functions $\nu_3(\tilde r), \nu_6(\tilde r)$ which are in (\ref{Usim3}). This means that the apparent freedom of these arbitrary functions  is not actual; hence there must be a coordinate transformation that can absorb these two arbitrary functions.

\section{Summary and Discussion}

We have shown how the method in \cite{edvic} which was used to investigate
conformally flat pure radiation spacetimes can be developed to
investigate the more complicated situation where there is a non-zero
cosmological constant; in particular,
we have found the subclass of conformally flat pure radiation spacetimes with negative cosmological constant   $\Lambda = - \tau\bar\tau $.

This analysis has extended our experience and knowledge of the GIF formalism, and in particular we have seen how in the GIF formalism we can handle spacetimes with multiple Killing vectors, by 'copying' tables from the generic case,  and also treat the one dimensional isotropy freedom of a null rotation.

As in \cite{edvic}, having constructed the spacetime via GIF, we find it is easy to read off the Karlhede classification; also as in \cite{edvic},  we needed to  go to the fourth order in the derivatives of the Riemann tensor, and, moreover, we were able to see directly how  different aspects of the Karlhede algorithm, especially regarding null isotropy, manifested themselves.  The fact that, for these two classes of spaces, we can carry out the Karlhede classification, {\it by hand} ,with  a simple calculation, emphasises the power of the GIF operators which we are able to use directly in place of the more complicated spinor calculations associated with the computer programmes for the Karlhede algorithm.  In fact in \cite{edvic} we could have simplified the Karlhede classification calculation, by changing from GIF operators to the simpler GHP scalar operators; this is permissable in \cite{edvic} because the second dyad spinor $\i(\equiv {\bf I})$, which enables us to translate from GHP formalism to GIF,  is intrinsic and invariant in the GIF.  On the contrary, for the spacetimes in this paper, we have seen that we do not get an intrinsic second spinor from the GIF formalism;  rather the spinor  ${\bf I \!\!\!\!\,* } $ which we use has one degree of freedom fixed in a non-intrinsic manner.  Therefore if we carry out a similar analysis as we did in the previous section {\it using GHP tables and operators}, we will not get a valid Karlhede classification: the analysis will not go any further than the second derivatives of $\Phi$ (essentially the GHP tables for $P$ and $Q$ (\ref{GHPtableP}), (\ref{GHPtableQ})).

In view of the complications and subtlties which arose for the spacetimes with zero cosmologiacal  constant, it will be interesting to see how the computer programmes handle these spacetimes, and especially the existence of one degree of null isotropy.

We have only given the discrete information regarding symmetries; using the method in \cite{edlud3} we will be able to find explicit expressions for the Killing vectors, and any homothetic Killing vectors present.

We have also used the GIF to construct the other spacetimes for this class --- those  with the condition $\Lambda \ne - \tau\bar\tau $; in this case there was no isotropy, but the calculations were longer and we will present the results elsewhere \cite{edram2}.  These calculations and results are enabling us  gradually to build up our experience and skill in the GIF,  so as to tackle even more complicated situations in the future.

Although there have recently been a number of investigations of pure radiation spacetimes with non-zero cosmological constant for different Petrov types of Weyl tensors \cite{czmc}, 
\cite{carm1}, \cite{carm2}, \cite{ozs},  \cite{bic1}, \cite{bic2}, \cite{pod}, \cite{gr1}, \cite{gr2}, these investigations generally seem to be built around a non-zero Weyl tensor, and it is not clear whether the whole class of conformally flat spaces are included as  special cases; moreover, the conformally flat limits do not seem to be easily deduced from the more general cases.
On the other hand, in this paper and in \cite{edram2} we have investigated the spacetimes with a formalism which is directly suited to the class, and the explicit metrics found here are in simple form. It remains to investigate the whole class of these spacetimes found via GIF, with the conformally flat limits of these various other investigations.

\section{Acknowledgements}

SBE wishes to thank Centro de Matem\'atica for supporting a visit to Universidade do Minho and the Mathematics Department for their hospitality.  MPMR wishes to thank Vetenskapsr\aa det (Swedish Research Council) for supporting a visit to Link\"opings universitet and the Mathematics Department for their hospitality.

\end{document}